\documentclass[superscriptaddress,amsmath,amssymb, aps, prx,longbibliography,nofootinbib,twocolumn]{revtex4-2}
\usepackage[normalem]{ulem}
\usepackage{float}
\usepackage{bm}
\usepackage{xcolor}
\usepackage[dvipsnames]{xcolor}
\usepackage{mathtools}  
\usepackage{graphicx}
\usepackage{hyperref}
\usepackage{dsfont}
\usepackage{xfrac}

\hypersetup{colorlinks,breaklinks,
            urlcolor=[rgb]{0.08,0.03,1},
            linkcolor=[rgb]{0.08,0.03,1},
            citecolor=[rgb]{0.08,0.03,1}}

\usepackage{booktabs} 

\makeatletter
\renewcommand\footnotesize{%
   \@setfontsize\footnotesize\@ixpt{11}%
   \abovedisplayskip 1\p@ \@plus2\p@ \@minus4\p@
   \abovedisplayshortskip \z@ \@plus\p@
   \belowdisplayshortskip 2\p@ \@plus2\p@ \@minus2\p@
   \def\@listi{\leftmargin\leftmargini
               \topsep 2\p@ \@plus2\p@ \@minus2\p@
               \parsep 1\p@ \@plus\p@ \@minus\p@
               \itemsep \parsep}%
   \belowdisplayskip \abovedisplayskip
}
\makeatother

\newcommand{\Harvard}{Department of Physics, Harvard University, Cambridge, Massachusetts 02138, USA.}
\newcommand{\TUM}{Technical University of Munich, TUM School of Natural Sciences, Physics Department, 85748 Garching, Germany}
\newcommand{\MCQST}{Munich Center for Quantum Science and Technology (MCQST), Schellingstr. 4, 80799 M{\"u}nchen, Germany}

\begin{document}


\title{Attraction from repulsion in an interacting Chern band: chiral superconductivity, generalized Hall crystals, and competing fractional Chern insulators}

\author{Fabian Pichler}
\thanks{These authors contributed equally to this work.\\}
\affiliation{\TUM}
\affiliation{\MCQST}
\author{Clemens Kuhlenkamp}
\thanks{These authors contributed equally to this work.\\}
\affiliation{\Harvard}
\author{Ashvin Vishwanath}
\affiliation{\Harvard}
\author{Michael Knap}
\affiliation{\TUM}
\affiliation{\MCQST}

\date{\today}

\begin{abstract}
Experiments realizing strong interactions in partially filled Chern bands have observed fractional Chern insulators (FCIs), superconductors, and topological crystals in close proximity.
Yet the microscopic relation between them remains poorly understood. 
Here, we use tensor network simulations to study repulsively interacting spinless fermions in a tunable topological band of the checkerboard lattice. 
We characterize the competing phases and identify key features of the experimental phenomenology. 
FCIs from the Jain sequence transition into chiral p-wave superconductivity both below and above half filling as the band dispersion is enhanced.
By contrast, at half filling, a generalized Hall crystal with quantized Hall conductance and a doubled unit cell forms for sufficiently strong coupling. 
Surprisingly, a weak-coupling approach starting from a metallic parent state captures the nature of both ordered phases at intermediate coupling. 
We find that fluctuations in the metal generate longer-range attraction, which we analyze within the random-phase approximation. 
This approach captures the pairing symmetry as well as the radial form factors of the Cooper pairs, and incorporating this attraction into Hartree-Fock selects the observed Hall crystal. 
Together, these results identify fluctuation-induced attraction from repulsion as a common microscopic ingredient of ordering, while our tensor network computations establish the direct competition of both phases with FCIs.
\end{abstract}

\maketitle


\section{Introduction}
\label{sec:Introduction}

Recent experiments in moir\'e materials have observed superconductivity of correlated electrons in topological bands, often near repulsively stabilized states such as fractional Chern insulators (FCIs) or electronic crystals~\cite{LongJuSc2025, xu_SC_2025,Kalantre2026}, suggesting a favorable setting for pairing from repulsion. In this context, it has been proposed that chiral superconductivity could emerge as a weak-coupling Kohn-Luttinger instability of a metal~\cite{Kohn1965, Luttinger1966}, further enhanced by quantum geometric effects~\cite{shavit2025QuantumGeometricKohnLuttinger,shavit2026NematicEnhancementSuperconductivity,jahin2026EnhancedKohnLuttinger, maymann2026HowPairingMechanism,luscherSuperconductivityChernBand2025,Xu2025_SC_theory,chen2026FinitemomentumSuperconductivityChiral,zou2026KohnLuttingerSuperconductivity}. Alternatively, strong-coupling `anyonic' descriptions of chiral superconductors~\cite{laughlinRelationshipHighTemperatureSuperconductivity1988, laughlinSuperconductingGroundState1988, fetterRandomphaseApproximationFractionalstatistics1989, leeAnyonSuperconductivityFractional1989, chenAnyonSuperconductivity1989} connected to nearby fractionalized phases have been put forward 
~\cite{Lee2018,Ma2020, Shi2024, divic_anyonsuperconductivity_2024, shi2025, pichlerMicroscopicMechanismAnyon,shiAnyonDelocalizationTransitions2025,Zhang2025, nosovAnyonSuperconductivityPlateau2025, shiNonAbelianTopologicalSuperconductivity2025,kuhlenkampRobustSuperconductivityDoping2025,chen_topological_2025,niuSimulatingTriangleHofstadterHubbard2025, ahnSuperconductivityProximateNonAbelian2025, gaoTopologicalCharge2neSuperconductors2025, hanExactModelsChiral2025, huangApparentInconsistencyStreda2025, nakajimaThermodynamicsDiluteAnyon2025,biborskiSignaturesSuperconductingPairing2025, seoUnifiedCategoricalDescription2026, vilardiSuperconductivitySpin2025, wangAnyonSuperfluidTrilayer2025, wangEmergentBosonic2025,  zhangCharge4eAnyonSuperconductor2025, zhangPathwaysChiralSuperconductor2025, zhouChernSimonsmatterConformalField2025, lotric2026phasesitinerantanyonslaughlins, zhang2026colorsuperconductorsholonmetals, fan2026hiddenweakpairingsuperconductivitynoninteracting, shi2026superconductivitynonfermiliquidmetals, wang2026topologicalsuperconductivityabelianfractional, senthil2026fractionalizedmetalsdopedanyons, han2026singlecomponenttwistedmathbbz3orthogonal, mehta2026coloringanyonsuperconductivity, hu2026quantumgeometryinduced}. While strong-coupling approaches often struggle to identify which phases are energetically favored by a given microscopic Hamiltonian, weak-coupling approaches are uncontrolled in the strongly correlated limit relevant to experiments. Identifying how repulsive interactions produce superconductivity and, within the same microscopic framework, how the superconducting state competes with nearby quantum Hall and crystalline phases remains an open problem. 

\begin{figure*}
\begin{center}
\includegraphics[width=0.98\linewidth]{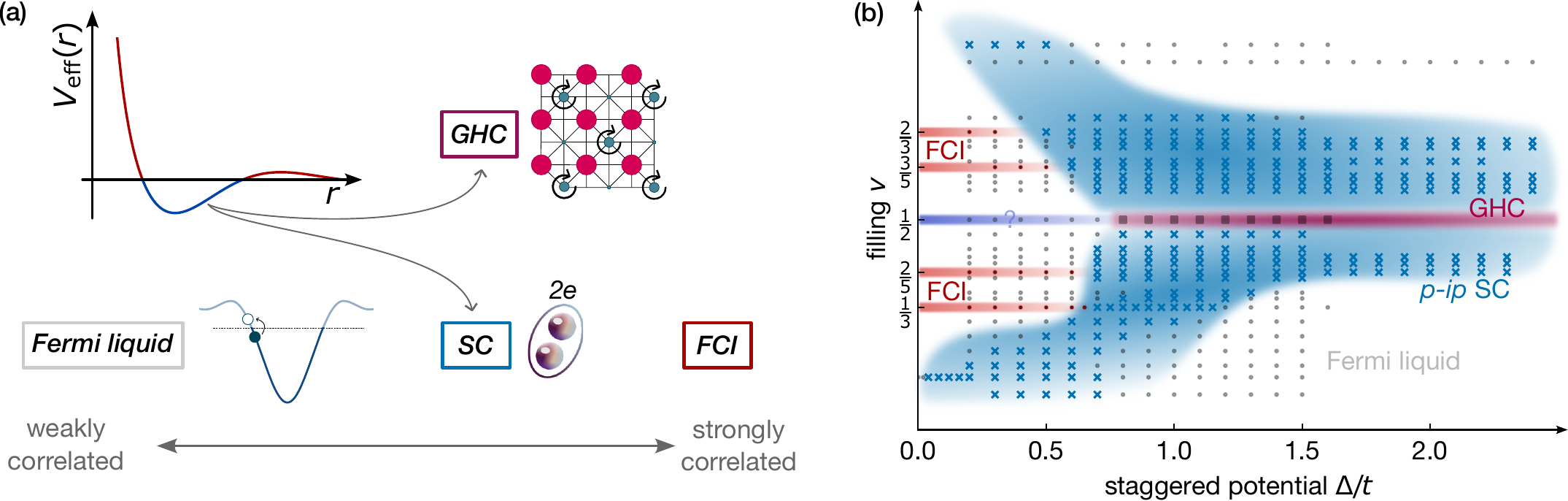}
\caption{\textbf{Mechanism and phase diagram.} \textbf{(a)} Higher-order particle-hole processes generate attractive longer-range components of the effective interactions starting from bare repulsion. This attraction favors both chiral superconductors (SC) and commensurate generalized Hall crystals (GHC). In the strongly correlated regime, the superconductor transitions into fractional Chern insulators (FCIs) for certain fillings. \textbf{(b)} Phase diagram of interacting spinless fermions in a Chern band, Eq.~\eqref{eq:model}, obtained from infinite density-matrix renormalization-group simulations on cylinders with $L_y = 8$ and bond dimension $\chi=1024$. The interaction strength is fixed at $V=2.2t$, while filling $\nu$ and the staggered potential $\Delta$ are tuned. Crosses correspond to ground states with the ratio of the charge-$2e$ and charge-$1e$ correlation lengths satisfying $\xi_{2e}/\xi_{1e}>1$, which we use as a diagnostic for signatures of superconductivity. Further analysis below identifies this phase as a $p-ip$ superconductor. At several filling fractions, FCIs are the ground states for small $\Delta$, transitioning into the SC for increasing $\Delta$. At $\nu=1/2$, a GHC is stabilized at intermediate $\Delta$. }
\label{fig:schematicPD}
\end{center}
\end{figure*}

Here, we study this interplay in a minimal setting: spinless fermions with repulsive interactions in a partially filled Chern band. Starting from a metallic parent state, we find that screening generates further-range attractive interactions within the random-phase approximation. At generic fillings, this attraction favors chiral superconductivity, while at half filling, lattice commensurability redirects the instability toward charge-density-wave order. Remarkably, this analysis suggests that the same longer-range attraction that induces superconductivity also stabilizes a commensurate generalized Hall crystal (GHC), a topological charge density wave state with a chiral edge mode, identifying a common microscopic ingredient linking the two phases (Fig.~\ref{fig:schematicPD}\textbf{a}). 

In the intermediate to strong-coupling regime, we uncover a rich phase diagram using unbiased tensor-network methods. We find that the system features a stable chiral $p$-wave superconductor (SC), as well as a GHC, fully consistent with the analytical predictions. Strikingly, the chiral superconductor gives way to FCIs at several Jain fillings as the bandwidth is reduced
(Fig.~\ref{fig:schematicPD}\textbf{b}).

Our findings identify proximity to a tunable van Hove singularity in repulsively interacting Chern bands as a key ingredient for stable chiral superconductivity. Strong signatures of the superconductor, which on the weak-coupling side can be explained through a Kohn-Luttinger-inspired attraction from repulsion mechanism, persist even in our finite-size numerics, extending continuously into the strongly interacting regime where the superconductor competes with fractional quantum Hall states.

The remainder of this work is organized as follows. In Sec.~\ref{sec:model}, we introduce the interacting lattice model and summarize its filling-dependent phase diagram. In Sec.~\ref{sec:weakcoupling}, we analyze the metallic regime within the random phase approximation and show that higher-order many-body processes generate longer-range attractive components of the effective interaction that favor chiral $p$-wave pairing. Section~\ref{sec:filling} presents our tensor-network results across a broad range of fillings and characterizes the resulting superconducting phase through its correlation lengths and Cooper-pair wave function. Additionally, we analyze the finite-circumference scaling toward two-dimensional superconducting order. In Sec.~\ref{sec:onethird}, we examine the phase diagram at $\nu=1/3$ in detail, including the bandwidth-tuned transition between the fractional Chern insulator and the superconductor. Section~\ref{sec:GHC} discusses the generalized Hall crystal found at half filling and relates its stability to the same effective attraction that promotes superconductivity. We conclude in Sec.~\ref{sec:Outlook} with implications for interacting Chern-band materials and open questions. Additional numerical results and parameter scans are
provided in the Appendices.

\section{Model and overview of the phase diagram}
\label{sec:model}

\begin{figure}
\begin{center}
\includegraphics[width=0.99\linewidth]{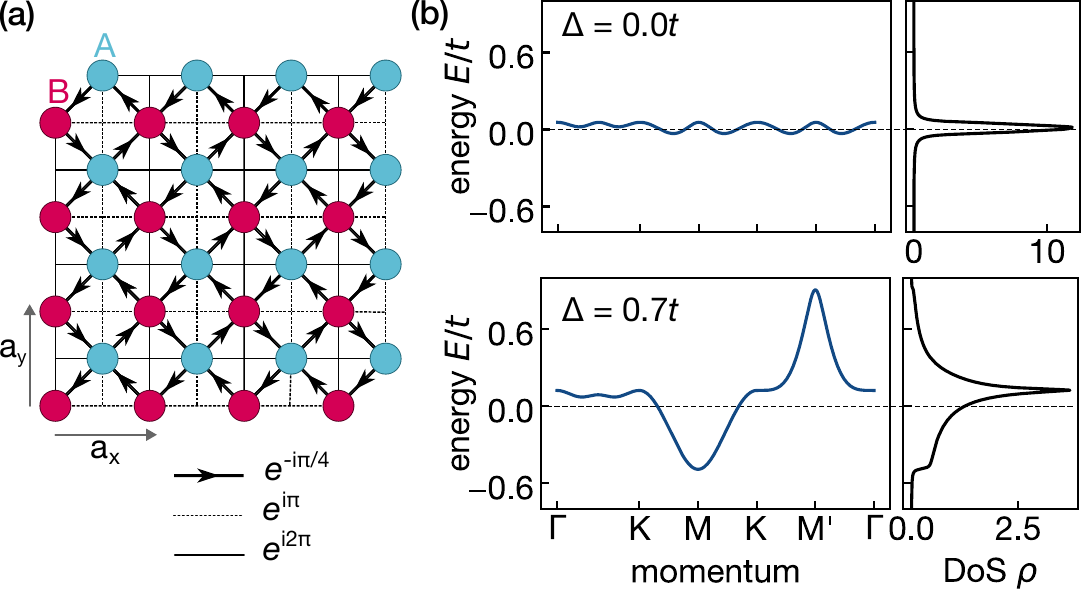}
\caption{\textbf{Model.} \textbf{(a)} Topological checkerboard model represented using a two-site unit cell containing sublattices $A$ and $B$. In the equivalent Hofstadter representation, each elementary plaquette carries $\pi$ flux, while the two-site cell encloses a total flux of $2\pi\equiv0$.
\textbf{(b)} Single-particle dispersion and corresponding density of states (DoS) of the lower band, with Chern number $C=1$, for staggered potential $\Delta=0$ (top) and $\Delta=0.7t$ (bottom). The dashed line corresponds to $\nu=1/3$ filling. At finite $\Delta$, there is a single Fermi surface. For small fillings below the van Hove singularity, it is particle-like and centered at the $M=(\pi, 0)$ point. For large fillings, it is hole-like and centered at $M'=(0, \pi)$. The other high-symmetry points of the Brillouin zone are $\Gamma=(0, 0)$ and $K=(\pi, \pi)$.  }
\label{fig:model}
\end{center}
\end{figure}

Chern bands with strong interactions offer a promising platform to realize a multitude of strongly correlated phases and their transitions. 
An open question, however, is which microscopic ingredients allow these phases to be realized within the same model. At first sight, their requirements appear to be in tension: stable FQH states favor a narrow Chern band dominated by strong repulsive interactions, whereas superconductivity requires an effective attractive pairing channel.
One might therefore envision tuning a transition into the superconducting phase by increasing the bandwidth. 
From the metallic side, the same transition can be viewed as arising from a Fermi surface on top of which attractive interactions are generated by higher-order many-body processes via a Kohn-Luttinger mechanism~\cite{Kohn1965,Luttinger1966}.
In this case, it is additionally favorable to retain a high density of states and a large quantum metric close to the Fermi surface~\cite{shavit2025QuantumGeometricKohnLuttinger,shavit2026NematicEnhancementSuperconductivity,jahin2026EnhancedKohnLuttinger, maymann2026HowPairingMechanism,luscherSuperconductivityChernBand2025,Xu2025_SC_theory,chen2026FinitemomentumSuperconductivityChiral,zou2026KohnLuttingerSuperconductivity}.
As a minimal example that incorporates these features, we consider a topological checkerboard lattice model of spin-polarized fermions~\cite{sun2011NearlyFlatbandsNontrivial, neupert2011, Sheng2011, regnault2011FractionalChernInsulator}
\begin{equation}
    H = H_0  + H_\Delta + V \displaystyle\sum_{i}\sum_{\boldsymbol{\delta}} n_{A}(\mathbf{x}_i) n_{B}(\mathbf{x}_i + \mathbf{\boldsymbol{\delta}})  \label{eq:model}
\end{equation}
with a kinetic term $H_0= \sum_{ij} t_{ij} c^\dagger_{i} c_{j}$ including up to third nearest-neighbor hopping $t_{ij}$, and nearest-neighbor density-density interactions $V$ (Fig.~\ref{fig:model}\textbf{a}). Here, $\boldsymbol{\delta}$ corresponds to the four nearest-neighbor vectors. 
We describe the model using a two-site unit cell containing sublattices $A$ and $B$, with complex inter-sublattice hopping and purely real intra-sublattice hopping, resulting in two Chern bands with Chern numbers $C=\pm1$. For vanishing next-nearest-neighbor hopping $t' = 0$, the single-particle model features two Dirac cones, at the $M=(\pi, 0)$ and $M'=(0, \pi)$ points of the Brillouin zone. A finite $t'$ instead opens up a topological gap in the single-particle bands.
The model can equivalently be viewed as a Hofstadter model with $\pi$ flux per elementary plaquette or as a two-sublattice Chern insulator with total flux $2\pi\equiv0$ through the two-site cell.
Unless stated otherwise, we choose the single-particle hopping parameters to minimize the bandwidth of the lower band: nearest, next-nearest, and third-nearest-neighbor hopping, respectively given by $t$, $t'= t/{(2+\sqrt2)}$, and $t''= t/{(2+2\sqrt2})$~\cite{sun2011NearlyFlatbandsNontrivial}. Additionally, we introduced in Eq.~\eqref{eq:model} a staggered potential
\begin{equation}
    H_{\Delta} = \Delta \sum_{i}[n_A(\mathbf{x}_i) - n_B(\mathbf{x}_i)],
\end{equation}
allowing us to tune the single-particle bandwidth (Fig.~\ref{fig:model}\textbf{b}). 
At $\Delta=0$, the Hamiltonian respects a one-site magnetic translation that exchanges the two sublattices, so that the primitive physical unit cell carries $\pi$ background flux. A finite $\Delta$ breaks this symmetry and makes the two-site $A$--$B$ cell the primitive unit cell, which encloses a total flux of $2\pi\equiv0$.
For vanishing $\Delta$, the model has been shown to host FCIs~\cite{neupert2011, Sheng2011,regnault2011FractionalChernInsulator}. We confirm the FCIs at the Jain fillings $\nu \in \{1/3, \; 2/5, \; 3/5, \; 2/3\}$ 
using infinite density-matrix renormalization-group (iDMRG) simulations~\cite{white1992DMRG, Hauschild2018Tenpy} and demonstrate that they are stable for a finite range of small $\Delta$ (see Appendix~\ref{app:FCI}). 

In the opposite limit of $\Delta \rightarrow \infty$, one sublattice is effectively projected out, thereby also eliminating the interaction term, which couples different sublattices. Consequently, in this limit, the model realizes a non-interacting topologically trivial square lattice, which results in a Fermi liquid (FL) for all partial fillings of the lower band. 
Hence, as the staggered potential is increased, the ground state at the Jain fillings must evolve from an FCI to a Fermi liquid. Although a direct first-order transition is conceivable, we find that the system instead forms a chiral $p-ip$ superconductor as an intermediate phase.
The compressible intermediate SC extends over a wide filling regime, remaining stable even away from the special filling factors of the Jain states. 
Among the fillings studied, $\nu=1/2$ is a notable exception. Here, a charge density wave is commensurate with the lattice and the superconductor is disfavored compared to a generalized Hall crystal, i.e., a charge density wave that doubles the unit cell with Chern number $C=1$. {Interestingly, this phase persists even when $\Delta$ tunes the system into the topologically trivial band.} 

A schematic overview of the phase diagram, obtained using iDMRG, is presented in Fig.~\ref{fig:schematicPD}\textbf{b}. 
We find a multitude of phases that have also been observed in experiments of strongly interacting Chern bands~\cite{LongJuSc2025,xu_SC_2025}.

\section{Attraction from repulsion}
\label{sec:weakcoupling}
\begin{figure}
\begin{center}
\includegraphics[width=0.99\linewidth]{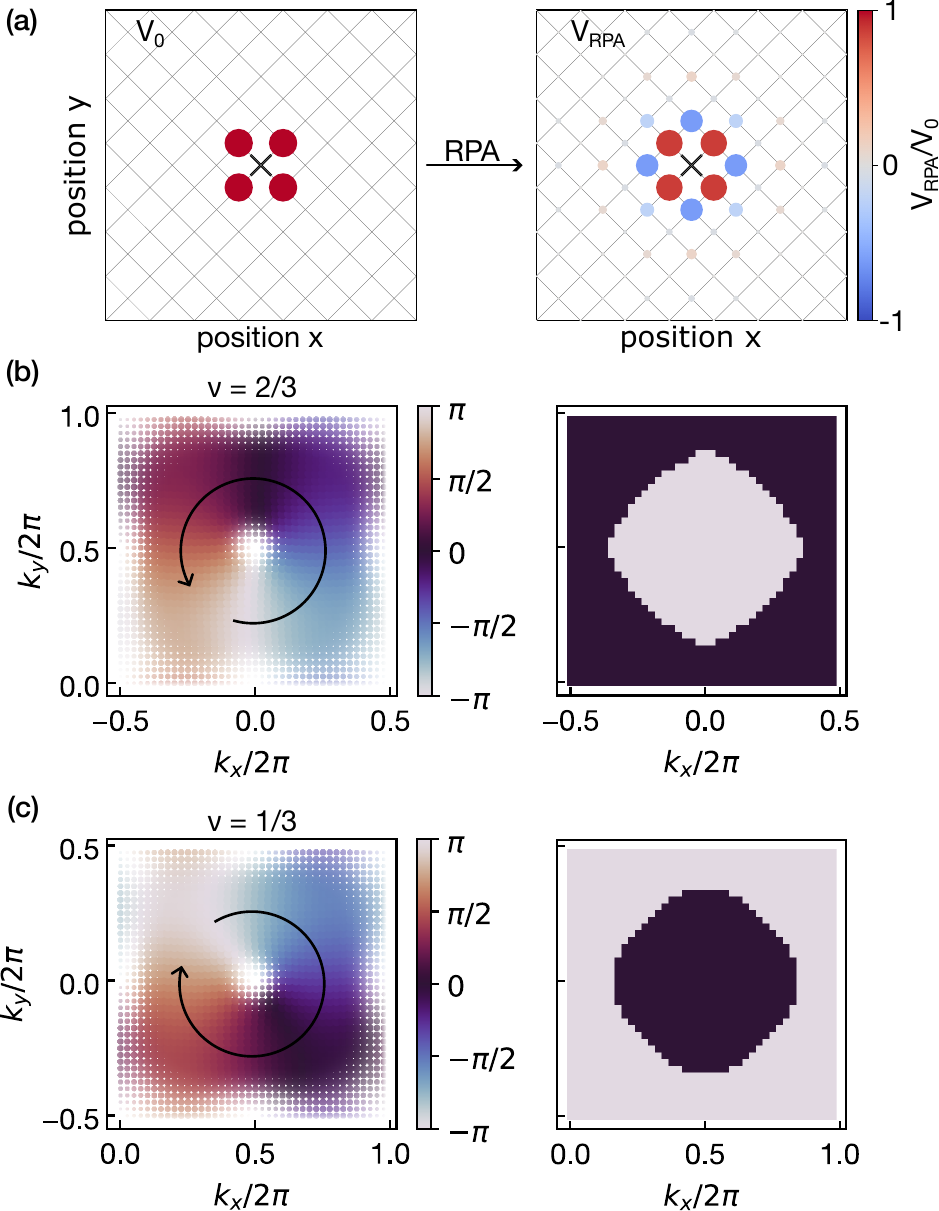}
\caption{\textbf{Effective attractive interactions and emergent superconducting pairing.} \textbf{(a)} The bare repulsive nearest-neighbor interactions (left) are dressed by higher order particle-hole corrections resummed by the random phase approximation (right). While the renormalized nearest-neighbor interaction remains repulsive, an attractive next-nearest-neighbor interaction is induced. $V_\mathrm{RPA}$ is shown here for $\nu=1/3$, $\Delta=0.9t$, $V=2.2t$.
 \textbf{(b, c)} Gap function obtained from self-consistently solving the superconducting gap equation (left) for two representative fillings and the corresponding Fermi surface (right) with filled states in black and empty states in white. The $\Delta_{BB}(\mathbf{k})$ component is shown. (\textbf{b}) For $\nu=2/3$, the Fermi surface is hole-like, while for $\nu=1/3$ (\textbf{c}), the Fermi surface is particle-like. In both cases, the gap function has a clear $p$-wave structure, but with opposite chirality. Since also the Fermi surfaces flip from particle- to hole-like, the BdG Chern number is $C_{\mathrm{BdG}}=-1$ in both cases, suggesting a $p-ip$ SC. }
\label{fig:KohnLuttinger}
\end{center}
\end{figure}

The appearance and robustness of the chiral superconductor, even in the strongly repulsive regime of our model, are the most striking features of the phase diagram.
Starting from a metallic state at large $\Delta$, higher-order particle-hole processes can generate attractive components of the effective interaction, as first demonstrated by Kohn and Luttinger~\cite{Kohn1965,Luttinger1966}. While this mechanism is inherently perturbative, it can be enhanced by a large density of states and nontrivial quantum geometry
~\cite{shavit2025QuantumGeometricKohnLuttinger,shavit2026NematicEnhancementSuperconductivity,jahin2026EnhancedKohnLuttinger, maymann2026HowPairingMechanism,luscherSuperconductivityChernBand2025,Xu2025_SC_theory,chen2026FinitemomentumSuperconductivityChiral,zou2026KohnLuttingerSuperconductivity}. To identify the relevant pairing channel in this weakly interacting limit, we use the random-phase-approximation (RPA) to determine the effective interaction profile.
For the chosen single-particle parameters, a small but finite staggered potential $\Delta$ produces a high density of states associated with a nearby van Hove singularity, which is expected to provide favorable conditions for superconductivity over a wide range of fillings; see Fig.~\ref{fig:model}\textbf{b}. 
To capture the screening contained in the higher-order many-body processes, we consider the renormalization of the bare interaction $V^{0}_\mathbf{q}$ within RPA:
 \begin{equation}
     V_\mathbf{q}^\mathrm{RPA} = V^{0}_\mathbf{q} \big[ 1 -  \Pi(\mathbf{q}) V_\mathbf{q}^0\big]^{{-1}} \label{eq:RPA_interaction}
 \end{equation}
which is understood as a matrix equation in sublattice space. We have introduced the static polarization bubble
\begin{equation}
    \Pi_{ab}(\mathbf{q}) = \frac{1}{N^{\text{\tiny UC}}} \sum_{\mathbf{k}, \mu\nu}\frac{n_{\mathbf{k}+\mathbf{q}}^\mu - n_{\mathbf{k}}^\nu}{\varepsilon_{\mathbf{k}+\mathbf{q}}^\mu - \varepsilon_{\mathbf{k}}^\nu} \Lambda_{\mathbf{k}+\mathbf{q}, \mathbf{k}}^{
    \mu \nu, a}(\Lambda_{\mathbf{k}+\mathbf{q}, \mathbf{k}}^{
    \mu \nu, b})^* .\label{eq:Pi_bubble}
\end{equation}
Here, $a,b\in\{A,B\}$ are sublattice indices, $\mu,\nu$ label the two bands, and $n_{\mathbf{k}}^\nu=f_\mathrm{FD}(\varepsilon_{\mathbf{k}}^\nu-\mu)$ is the Fermi occupation.
The bare repulsive $V^{0}_\mathbf{q}$ is obtained by Fourier transforming the nearest-neighbor interactions of Eq.~\eqref{eq:model}
\begin{equation}
    H_\mathrm{int} = \frac{1}{N^{\text{\tiny UC}}} \sum_{ab} \sum_{\mathbf{q} \mathbf{k} \mathbf{k}'} V^0_{ab}(\mathbf{q}) c^\dagger_{a, \mathbf{k}-\mathbf{q}} c^\dagger_{b, \mathbf{k}'+\mathbf{q}} c_{b, \mathbf{k}'} c_{a, \mathbf{k}}. \label{eq:interaction_full_ab}
\end{equation}
The form factors $\Lambda_{\mathbf{k}, \mathbf{k}'}^{\mu \nu, a} = \mathcal{U}_{\mu a}^*(\mathbf{k}) \mathcal{U}_{\nu a}(\mathbf{k}')$ are defined in terms of the band eigenstates $\mathcal{U}_{\mu a}(\mathbf{k})$ associated with the energy eigenvalues $\varepsilon_{\mathbf{k}}^\mu$ of the single-particle Hamiltonian. To improve convergence, we calculate Eq.~\eqref{eq:Pi_bubble} at a finite but small temperature of $T=10^{-3} t$, working with a uniform momentum grid containing $L^2 = 1600$ points in the first Brillouin zone. We find that the renormalized interactions remain repulsive on nearest-neighbor sites but become attractive for next-nearest neighbors, which are the closest intra-sublattice interactions; see Fig.~\ref{fig:KohnLuttinger}\textbf{a}. 

We now restrict the renormalized effective interactions to the Cooper channel at zero center-of-mass momentum
\begin{equation}
    H_\mathrm{int} =  \frac{1}{N^{\text{\tiny UC}}} \sum_{\substack{ab \\ cd}} \sum_{\mathbf{p}, \mathbf{k}} V_{ab, cd}(\mathbf{p}, \mathbf{k}) c^\dagger_{a, \mathbf{p}} c^\dagger_{b, -\mathbf{p}}
    c_{c, -\mathbf{k}} c_{d, \mathbf{k}}
\end{equation}
with $V_{ab, cd}(\mathbf{p}, \mathbf{k}) = V^\mathrm{RPA}_{ab}(\mathbf{k} - \mathbf{p} ) \delta_{ad} \delta_{bc}$. We then define the Cooper pair gap function as
\begin{equation}
    \Delta_{ab}(\mathbf{p}) = - \frac{1}{N^{\text{\tiny UC}}} \sum_{\substack{\mathbf{k}\\ cd}} \mathcal{W}_{ab, cd}(\mathbf{p}, \mathbf{k}) \langle c_{c, -\mathbf{k}} c_{d, \mathbf{k}}\rangle,
\end{equation}
where we have introduced the explicitly antisymmetrized Cooper-pair vertex, consistent with fermionic exchange statistics: 
\begin{equation}
    \mathcal{W}_{ab, cd}(\mathbf{p}, \mathbf{k}) = \frac{1}{2} \big[V_{ab, cd}(\mathbf{p}, \mathbf{k})  - V_{ab, dc}(\mathbf{p}, -\mathbf{k}) \big].
\end{equation}

We solve for the superconducting gap equation self-consistently by iteratively diagonalizing the BdG Hamiltonian, which in the Nambu basis
$\Psi_{\mathbf{k}}=(c_{\mathbf{k}A},c_{\mathbf{k}B},
c^\dagger_{-\mathbf{k}A},c^\dagger_{-\mathbf{k}B})^T$ is
\begin{equation}
H_{\mathrm{BdG}}(\mathbf{k})=
\begin{pmatrix}
h(\mathbf{k})-\mu & \Delta(\mathbf{k})\\
\Delta^\dagger(\mathbf{k}) & -h^T(-\mathbf{k})+\mu
\end{pmatrix}.
\end{equation}
We find that inter-band contributions are essential for obtaining a robust pairing solution.
The obtained gap function has a chiral $p$-wave symmetry; see Fig.~\ref{fig:KohnLuttinger}\textbf{b, c}. 
Across the range of fillings studied, the converged BdG spectrum is gapped and has Chern number $C_{\mathrm{BdG}}=-1$, corresponding to a chiral central charge $c_-=C_{\mathrm{BdG}}/2=-1/2$.
Thus, the BdG Chern number always opposes the chirality of the parent Chern band, as predicted for pairing-by-repulsion~\cite{maymann2026HowPairingMechanism}. The chirality of the pairing wave function does not fix the BdG Chern number, which additionally depends on whether the Fermi surface is particle- or hole-like. In our system, the Fermi surface changes from particle-like for small fillings to hole-like for large fillings, as the van Hove singularity is crossed. Remarkably, we find that the chirality of the pairing function changes concurrently, such that the BdG-Chern number, and hence the chiral central charge $c_- = -1/2$, remains filling-independent.
{This result is not captured by a band-projected treatment: obtaining the $\nu > 1/2$ state from the $\nu < 1/2$ results by a particle-hole transformation within the projected lower band would instead predict $c_- = 1 - (-1/2) = 3/2$. Including the remote band therefore qualitatively changes the topology of the paired state.}
Furthermore, the BdG-Chern number remains finite even for large $\Delta$, in a parameter regime where the parent band has become trivial (the single-particle gap closes at $\Delta \approx 1.1t$).  

\section{Chiral superconductivity in finite size numerics}
\label{sec:filling}
While Kohn-Luttinger superconductivity is a generic instability of a metal, pairing scales are typically strongly suppressed, which makes this effect challenging to observe. Here, we study our model using unbiased tensor-network simulations and demonstrate that the predicted superconductor robustly appears even in finite systems and in the presence of strong repulsion. 

We represent states as infinite matrix product states (iMPS) on semi-infinite cylinders, and find the ground state using iDMRG. We make use of $U_c(1)$ charge and $\mathbb{Z}_{L_y/2}$ momentum conservation along the cylinder circumference. If not stated otherwise, we consider cylinders with a circumference of $L_y=8$ sites, corresponding to $N_y^{\text{\tiny UC}} = L_y/2 = 4$ two-site A--B unit cells. 
For cylinders with an even number of unit cells, we impose antiperiodic boundary conditions by threading a flux $\Phi_y=\pi$. This shifts the allowed transverse momenta such that the Fermi surface is intersected by an even number of momentum cuts, avoiding an obstruction to pairing. This flux sector also has a lower energy than the periodic sector.
For each filling, we check increasingly larger $N_x^{\text{\tiny UC}}$ unit cell sizes along the infinite direction, to ensure that our results do not depend on the size of the chosen unit cell. 

To diagnose the emergence of superconducting order, we study the correlation length of operators in specific $U_c(1)$ charge and $\mathbb{Z}_{L_y/2}$ momentum sectors, which we label by $(Q, k_y)$. 
For an iMPS, the correlation length can be extracted from the second-largest eigenvalue $\lambda_{Q, k_y}^{(1)}$ of the transfer matrix in the corresponding charge sector~\cite{Hauschild2018Tenpy}
\begin{equation}
    \xi_{Q, k_y} = - \frac{1}{\log \lambda_{Q, k_y}^{(1)}}.
\end{equation}
We define $\xi_Q \equiv \max_{k_y}[ \xi_{Q, k_y}]$.
For fully gapped phases, such as FCIs, all correlation lengths quickly converge to a small finite value with increasing bond dimension $\chi$, since $\xi_{Q}$ is determined by the gap in the charge sector $Q$. 
In contrast, in a superconductor, the charge-$2e$ sector exhibits long-range correlations, while the $1e$ sector remains gapped. Thus, the charge-$2e$ correlation length must diverge $\xi_{2e}\rightarrow \infty$ while the charge-$1e$ correlation length remains finite $\xi_{1e}\rightarrow \text{const.}$ as $\chi \rightarrow \infty$. 
Hence, for a superconductor, the ratio $\xi_{2e}/\xi_{1e}$ diverges with bond dimension, while it saturates for competing insulators and metals.

We first discuss general trends as a function of the filling $\nu$ and the staggered potential $\Delta$, for a fixed interaction $V=2.2t$. We observe superconductivity for a wide range of fillings and $\Delta$'s, which we identify by a large and growing ratio $\xi_{2e}/\xi_{1e}$ (see Fig.~\ref{fig:schematicPD}\textbf{b}, Appendix~\ref{app:fillingPhaseDiagram}). 
Remarkably, we find no signatures of superconductivity at $\nu=1/2$, where a charge density wave is commensurate with the lattice. Instead, a symmetry-broken generalized Hall crystal with $\sigma_{xy}=1$ (in units of $e^2/h=1$) is stabilized for intermediate $\Delta$. We discuss this phase in Sec.~\ref{sec:GHC}.

To characterize the superconductor, we analyze the Cooper-pair wave function 
\begin{equation}
    \Psi_{ab}(\mathbf{x}) = \langle c_{a}(\mathbf{x}) e^{i \theta_{ab}(\mathbf{x})}c_{b}(\mathbf{0}) \rangle  \label{eq:pairwavefunction}
\end{equation}
where the phases $\theta_{ab}(\mathbf{x}) = \int_{\mathbf{r}_b}^{\mathbf{x}+\mathbf{r}_a} d\mathbf{r} \cdot \mathbf{A}(\mathbf{r})$ ensure that the Cooper-pair wave function transforms properly under $U(1)$ gauge transformation of the external gauge field $\mathbf{A}(\mathbf{x})\rightarrow\mathbf{A}(\mathbf{x})+\nabla\lambda(\mathbf{x})$. Here $\mathbf{r}_a$ is the position of sublattice $a \in \{A, B\}$ within the unit cell. We work in a fixed charge sector, so $\langle\Psi_{ab}(\mathbf{x}) \rangle = 0$ by symmetry.
 We instead evaluate~\cite{Sahay23, kuhlenkampRobustSuperconductivityDoping2025} 
\begin{equation}
\begin{aligned}
    C^\text{pair}_{ab}(\mathbf{x}; \mathbf{r}_0) &= \langle \Delta_\mathrm{ref}(\mathbf{r}_0) c_{a}(\mathbf{x}) e^{i \theta_{ab}(\mathbf{x})}c_{b}(\mathbf{0}) \rangle \\
    &\approx \alpha_\mathrm{ref}(\mathbf{r}_0) \Psi_{ab}(\mathbf{x})
\end{aligned}
\end{equation}
with $\Delta_{\mathrm{ref}}(\mathbf{r}_0) = c^\dagger_{B k_1}(\mathbf{r}_0) c^\dagger_{B k_2}(\mathbf{r}_0)$ at some reference site $|\mathbf{r}_0| \gg \ell_\mathrm{CP}$. Here, $\ell_\mathrm{CP}$ is the spatial extent of the Cooper pair and $k_1=-k_2$ are momenta along the periodic $y$ direction, corresponding to the momentum sector with the largest $\xi_{2e}$ correlation length. We have verified that as long as $\xi_{2e}$ and $|\mathbf{r}_0|$ are sufficiently large compared to $\ell_{\mathrm{CP}}$, the correlation function $C^\text{pair}_{ab}(\mathbf{x}; \mathbf{r}_0)$ does not depend on $\mathbf{r}_0$ up to an overall constant $\alpha_\mathrm{ref}(\mathbf{r}_0) \in \mathbb{C}$.

\begin{figure}
\begin{center}
\includegraphics[width=0.99\linewidth]{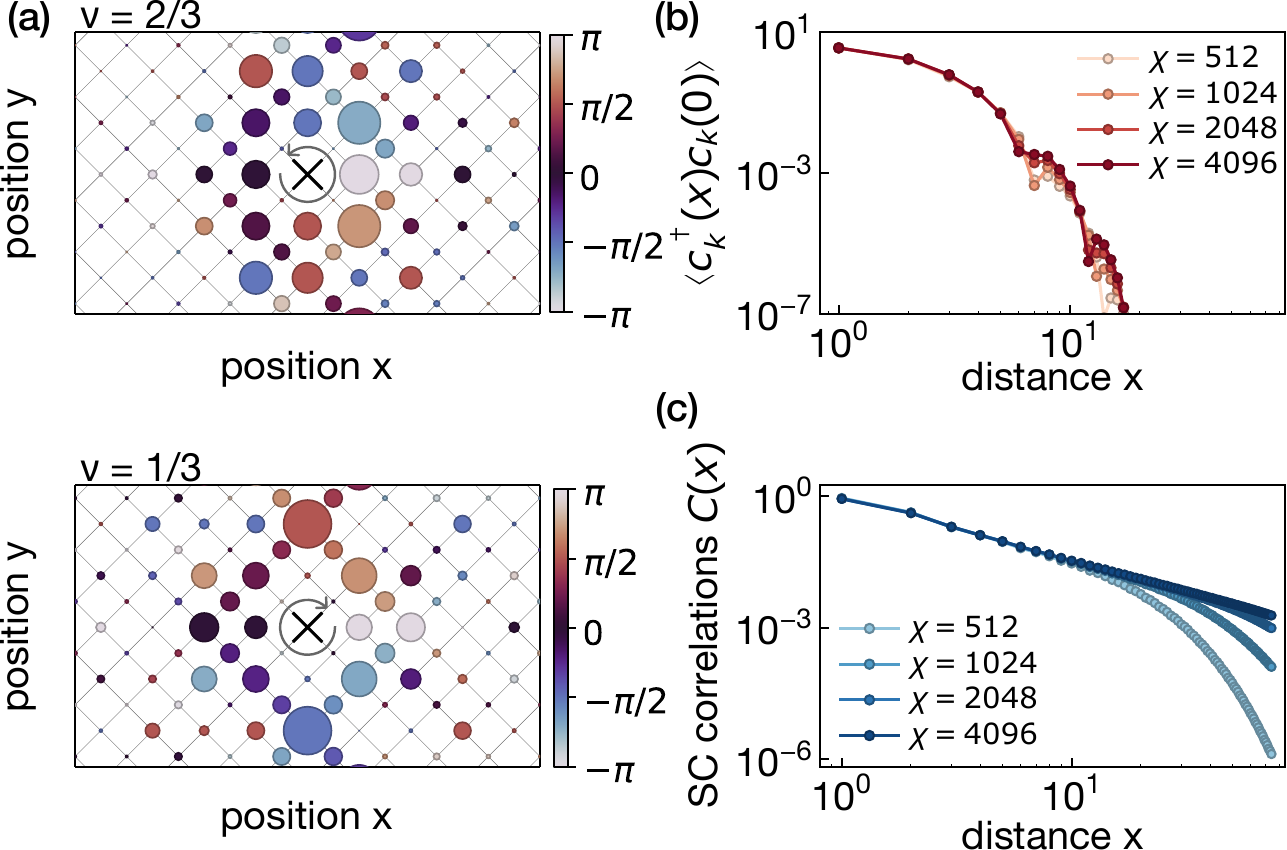}
\caption{\textbf{Cooper-pair wave function and correlations.} \textbf{(a)} Real-space pairing function Eq.~\eqref{eq:pairwavefunction} from iDMRG for $\nu=2/3$ (top) and $\nu=1/3$ (bottom). In both cases, the origin is placed on a $B$ sublattice site, and $\Delta=0.9t$, $V=2.2t$ were used.
The phase winds $|2\pi|$ around the origin, with opposite chirality in the two cases, consistent with our weak-coupling results (see Fig.~\ref{fig:KohnLuttinger}). \textbf{(b, c)} Correlation functions inside the SC phase ($\Delta=0.9t$, $V=2.2t$, $\nu=1/3$). Single-particle correlations \textbf{(b)} decay exponentially and quickly converge with bond dimension, while Cooper-pair correlations Eq.~\eqref{eq:SC_correlation_function} \textbf{(c)} decay algebraically, with the algebraic regime being limited only by bond dimension.}
\label{fig:pairing_WF_DMRG}
\end{center}
\end{figure}

We find that deep inside the SC phase, the Cooper-pair wave function is tightly localized, except for a nearly vanishing amplitude on nearest-neighbor sites (see Fig.~\ref{fig:pairing_WF_DMRG}\textbf{a} and Appendix~\ref{app:Cooperpair_wavefunction}). 
This is suggestive of effective repulsive interactions between nearest neighbors and attractive interactions beyond, in excellent agreement with the RPA-renormalized interaction discussed in Sec.~\ref{sec:weakcoupling}. 
Due to the presence of a finite staggered potential $\Delta$, the amplitude of the wave function is generally larger when the origin is placed on a $B$ site where the potential is smaller.
The Cooper-pair wave function is odd under inversion, and we find that it winds with a phase of $|2\pi|$ around the origin, suggesting a $p$-wave superconductor. This is in contrast to the SC observed in Refs.~\cite{wangChiralSuperconductivityFractional2025, guerciFractionalizationChiralTopological2025, guerci2026topologicalSC}, where a phase winding of $|6\pi|$ was found, but is consistent with our weak-coupling results (see Sec.~\ref{sec:weakcoupling} and Fig.~\ref{fig:KohnLuttinger}\textbf{b,c}), with the same reversal of the chirality when tuning the density through $\nu=1/2$.

\begin{figure}
\begin{center}
\includegraphics[width=0.99\linewidth]{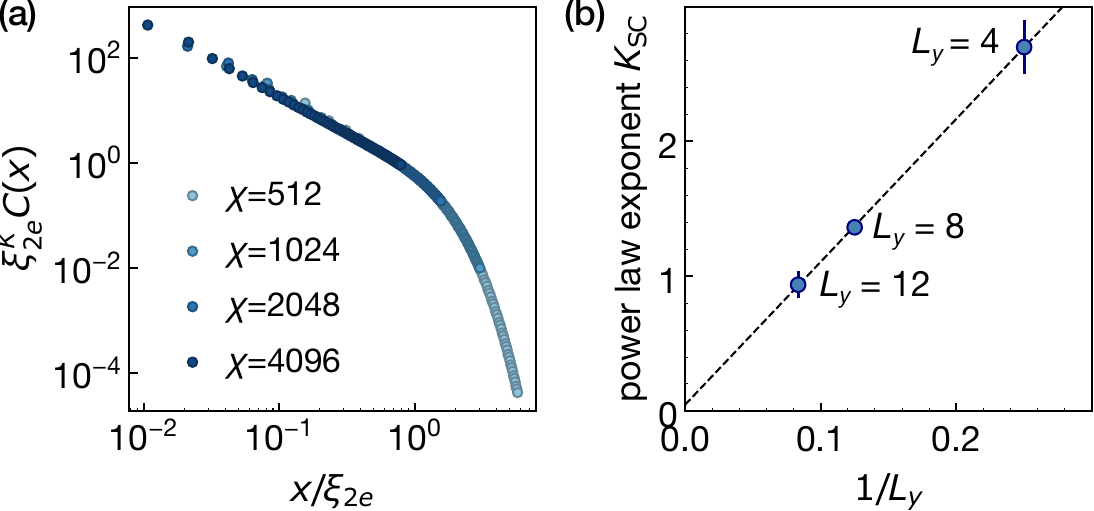}
\caption{\textbf{Power law exponent and system size scaling.} \textbf{(a)} Scaling collapse of Cooper-pair correlations with bond dimension for $L_y=8$ and $\nu=1/3$, resulting in a power-law exponent of $K_\mathrm{SC}=1.36\pm0.02$. \textbf{(b)} Evolution of the power-law exponent $K_\mathrm{SC}$ as a function of the (inverse) cylinder circumference. A trend towards decreasing $K_\mathrm{SC}$ with increasing $L_y$ is observed. Within error bars, $K_\mathrm{SC}$ extrapolates to zero for $L_y\rightarrow\infty$, consistent with the emergence of long-range order in the two-dimensional limit.}
\label{fig:scalingcollapse}
\end{center}
\end{figure}

Due to the quasi-one-dimensional geometry, the system cannot develop true superconducting long-range order. Instead, the intermediate phase realizes a Luther-Emery liquid with algebraically decaying superconducting correlations~\cite{Gannot2023} while also developing a single-electron gap. We characterize the SC correlations through
\begin{equation}
    C_{ab}(x;k_1,k_2)
    =
    \langle
    c^\dagger_{a k_1}(x)c^\dagger_{b k_2}(x)
    c_{b k_2}(0)c_{a k_1}(0)
    \rangle ,
    \label{eq:SC_correlation_function}
\end{equation}
where $a,b\in\{A,B\}$ denote sublattices and $k_1,k_2$ are transverse momenta (see Fig.~\ref{fig:pairing_WF_DMRG}\textbf{c}). In the dominant pairing channel, $a=b=B$ and
$k_1=-k_2=\pi/4$, the correlations decay as
\begin{equation}
    C_{BB}(x;\pi/4,-\pi/4)\sim x^{-K_{\mathrm{SC}}},
\end{equation}
which defines the superconducting power-law exponent
$K_{\mathrm{SC}}$. For a finite bond dimension $\chi$, this algebraic decay is cut off by the finite-entanglement correlation length $\xi_{2e}(\chi)$. We therefore extract $K_{\mathrm{SC}}$ from the scaling form
\begin{equation}
    C_{BB}(x;\chi)
    =
    \xi_{2e}(\chi)^{-K_{\mathrm{SC}}}
    f\!\left(\frac{x}{\xi_{2e}(\chi)}\right),
\end{equation}
following Refs.~\cite{Sahay23,kuhlenkampRobustSuperconductivityDoping2025}.
We obtain a scaling collapse with $K_{\mathrm{SC}}=1.36\pm0.02$ (Fig.~\ref{fig:scalingcollapse}\textbf{a}). Algebraic correlations are found only in the channel $k_1=-k_2=\pi/4$, while all other transverse-momentum channels decay exponentially. This is consistent with pairing at zero center-of-mass momentum. As expected for a chiral $p$-wave superconductor with a gapped charge-$1e$ sector, all single-particle correlations $\langle c_{a k}^\dagger(x) \, c_{b k}(0)\rangle$ decay exponentially and quickly
converge with bond dimension (see Fig.~\ref{fig:pairing_WF_DMRG}\textbf{b}).

\begin{figure*}
\begin{center}
\includegraphics[width=0.99\linewidth]{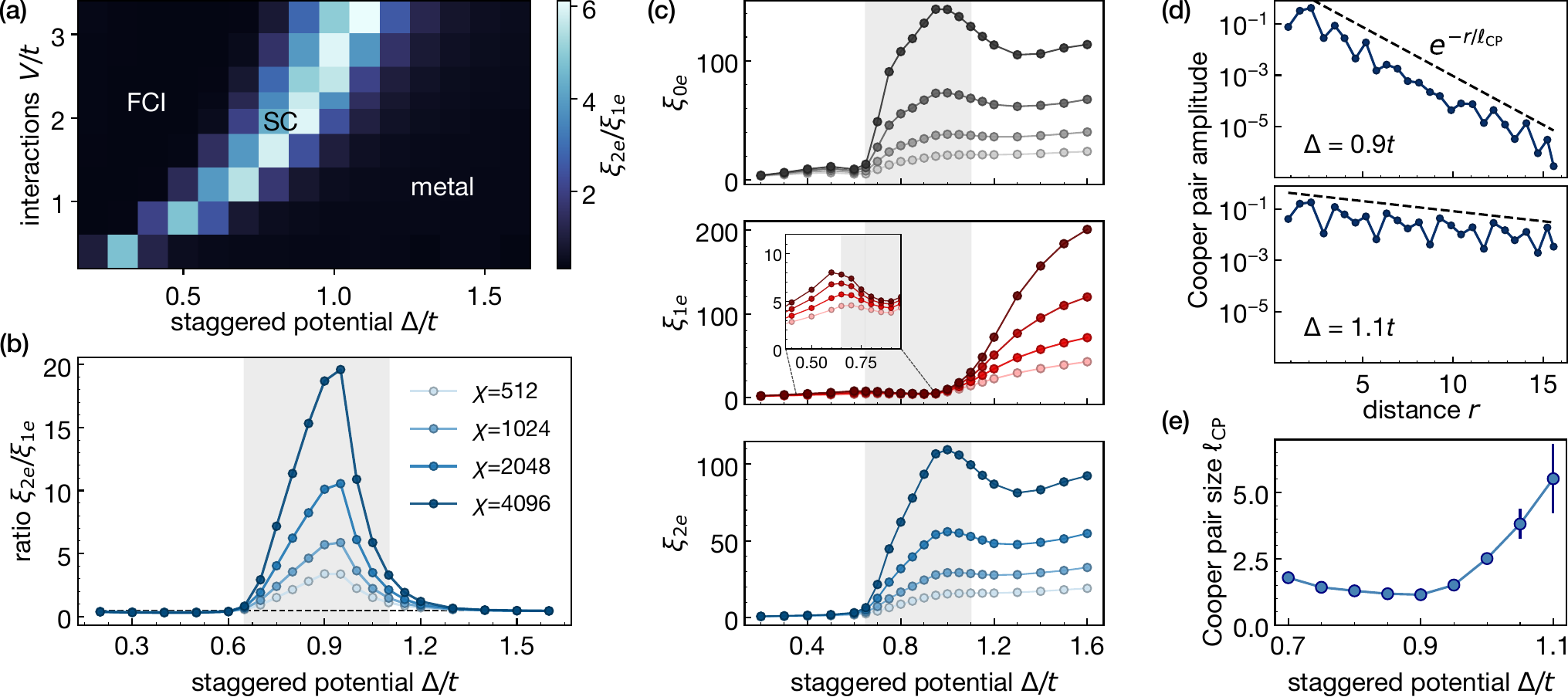}
\caption{\textbf{Bandwidth-tuned transition between an FCI and superconductor at $\nu=1/3$ for $L_y=8$.} \textbf{(a)} Ratio of correlation lengths $\xi_{2e}/\xi_{1e}$ for fixed bond dimension $\chi=1024$ as function of interaction strength $V$ and staggered potential $\Delta$ at $\nu=1/3$. We find an FCI for small $\Delta$ and large $V$, which is separated from a metallic phase in the large $\Delta$ and small $V$ regime by an intermediate superconducting phase. \textbf{(b)} Ratio of correlation lengths $\xi_{2e}/\xi_{1e}$ for fixed $V=2.2t$ as a function of $\Delta$ and various bond dimensions $\chi$. Both in the FCI (small $\Delta \lesssim 0.6t$) and the metal (large $\Delta \gtrsim 1.2t$), the ratio quickly converges in bond dimension to $\xi_{2e}/\xi_{1e} =0.5$ (dashed line). In the intermediate regime, the system forms a superconductor, and the ratio takes much larger values and grows with bond dimension. \textbf{(c)} Correlation lengths in different charge sectors as a function of the staggered potential $\Delta$, for fixed $V=2.2t$.
 \textbf{(d)} Squared amplitude of the Cooper-pair wave function at $\nu=1/3$ as a function of the distance from the origin for two different $\Delta$. The dashed lines correspond to a fitted exponential decay with Cooper pair size $\ell_\mathrm{CP}$, which is shown as a function of $\Delta$ in panel \textbf{(e)}. As the transition to the Fermi liquid is approached with increasing $\Delta$, the Cooper pair becomes larger. The minimum of $\ell_\mathrm{CP}$ is deep inside the SC phase.}
\label{fig:correlations}
\end{center}
\end{figure*}

For superconducting long-range order to emerge in the two-dimensional limit, pair correlations must become longer-range with increasing $L_y$. We confirm this behavior by extracting $K_\text{SC}$ for $L_y=4,8,12$ (Fig.~\ref{fig:scalingcollapse}\textbf{b}), which extrapolates to zero as $1/L_y \rightarrow 0$.

We highlight that while the observed structure of the Cooper-pair wave function can be qualitatively explained from a weak-coupling RPA perspective, we observe signatures of superconductivity only for intermediate to strong-couplings (see Fig.~\ref{fig:schematicPD}\textbf{b} and Appendix~\ref{app:fillingPhaseDiagram}). Remarkably, the superconductor even persists all the way to the FCI regime, indicating that strong interactions play an important role in stabilizing the superconductor.

\section{Transition from SC to FCI at 1/3 filling} \label{sec:onethird}
Having established the pairing structure of the superconductor, we now focus on the phases that are stabilized in our model at a representative filling of $\nu=1/3$. 

In the strongly interacting regime, where the interaction strength $V$ is much larger than the bandwidth, and for sufficiently small staggering field $\Delta$, we find a stable FCI, which we identify through the characteristic entanglement spectrum and quantized charge pumping under flux insertion (see Appendix~\ref{app:FCI}). 
Deep inside the FCI, all correlation lengths are small and quickly converge with bond dimension $\chi$, as expected for a gapped state.
By contrast, in the weakly interacting regime, which corresponds to large $\Delta$ at fixed interactions $V$, we find large correlation lengths in all charge sectors, which continue to grow with increasing MPS bond dimension (see Fig.~\ref{fig:correlations}\textbf{c}).
In this limit, the largest correlation length is in the charge-$1e$ sector, and we find that the charge-$2e$ correlation length takes the trivial value $\xi_{2e} \approx \xi_{1e}/2$. This indicates the formation of a Fermi liquid, which connects to the non-interacting state formed when $\Delta\rightarrow \infty$ and $\nu<1$.

Remarkably, in an intermediate regime, the $1e$ correlations are small and quickly converge with bond dimension, while both the $2e$ correlation length $\xi_{2e}$ and the charge neutral correlation length $\xi_{0e}$ continue to grow with bond dimension, leading to ratios $\xi_{2e}/\xi_{1e} \gg 1$ (see Fig.~\ref{fig:correlations}\textbf{a-c}).
This evidences the formation of a superconductor with a charge-$1e$ gap, while the charge-$2e$ sector is gapless.
We remark that the growing charge-neutral correlations $\xi_{0e}$ for $L_y=8$ are expected for quasi-1d superconductors. We have checked that, with increasing $L_y$, they behave opposite to the SC correlations and become shorter range.

\begin{figure*}
\begin{center}
\includegraphics[width=0.99\linewidth]{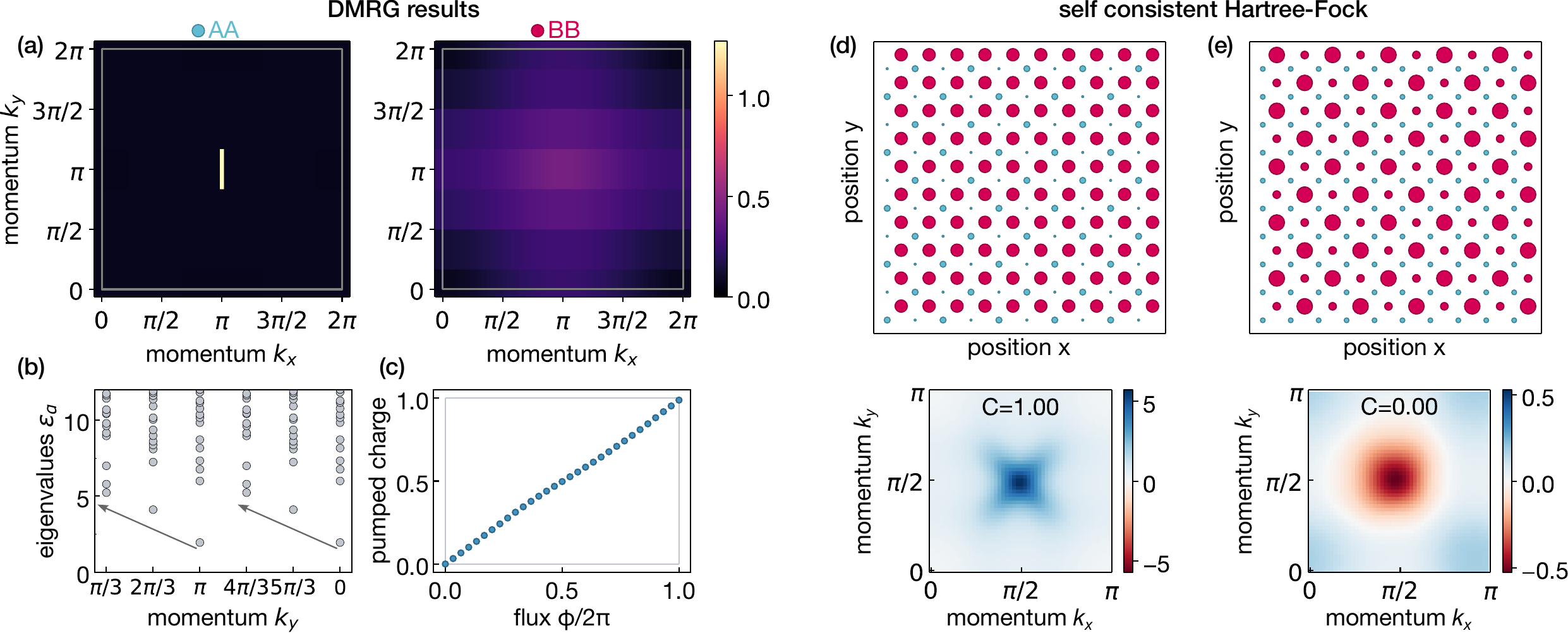}
\caption{\textbf{Generalized Hall crystal at half filling.} \textbf{(a)} Sublattice-resolved structure factor obtained from the iDMRG ground state for $L_y=12$, $\Delta=0.8t$ and $V=2.2t$ at $\nu=1/2$. On the minority sites (left panel, sublattice $A$, disfavored by the staggered potential), the structure factor has a peak at momentum $\mathbf{k}=(\pi, \pi)$, implying the formation of a CDW on that sublattice. By contrast, the structure factor for the $B$ sublattice (right panel) is comparatively uniform. \textbf{(b)} Entanglement spectrum showing a chiral edge mode with characteristic counting $(1, 1, 2, ...)$. Due to spontaneous breaking of translations, the mode folds back in momentum space, resulting in a replica. \textbf{(c)} Charge pumping: when inserting one flux quantum, a single charge is pumped. \textbf{(d, e)} Self-consistent Hartree-Fock calculations at $\Delta=0.8t$ with repulsive nearest-neighbor interactions $V=2.2t$ and (\textbf{d}) attractive or (\textbf{e}) repulsive next-nearest-neighbor interactions $V'=\mp 0.1t$. The top panel shows the resulting average density, while the bottom panel shows the Berry-curvature distribution of the filled band. In both cases, a CDW forms, opening up a gap at $\nu=1/2$. For attractive $V'<0$, the CDW forms on the minority sublattice and the resulting state has a finite Chern number $C=1$, which we refer to as the GHC. Conversely, for repulsive $V'>0$, the CDW forms on the majority sublattice, leading to a trivial state. }
\label{fig:halffilling}
\end{center}
\end{figure*}

We highlight that our finite cylinder numerics with $L_y=8$ are consistent with a continuous FCI to SC transition for $\nu=1/3$, although we cannot rule out that the transition may become (weakly) first order in the thermodynamic limit. Nonetheless, there are no clear discontinuities in the correlation lengths (see Fig.~\ref{fig:correlations}\textbf{c}) and both the energy and its first (numerical) derivative are continuous across the transition. Only its second derivative shows non-analytic behavior at the estimated critical point, consistent with a second-order transition (see Appendix~\ref{app:FCI}). Furthermore, quantized charge pumping, which we use to identify the FCI, survives up to $\Delta=0.8t$, overlapping with large $2e$ correlations indicative of the superconductor (see Appendix~\ref{app:FCI}). 
Note that while the $1e$ correlation length is small and quickly converges in bond dimension deep within the SC phase, it peaks at the transition point between the FCI and the SC and continues to grow with $\chi$ (see Fig.~\ref{fig:correlations}\textbf{c}). This indicates that at the critical point, both the $1e$ and $2e$ gap close~\cite{Ma2020}.
The numerics suggest that the $1e$ gap then reopens as the system is tuned into the SC, while the $2e$ gap remains closed.

This picture is further confirmed by analyzing the Cooper-pair size $\ell_\mathrm{CP}$, which we define by fitting an exponential decay $\sim e^{-r/\ell_{\mathrm{CP}}}$ to the squared amplitude of the Cooper-pair wave function Eq.~\eqref{eq:pairwavefunction}. It has a local minimum deep inside the superconducting phase at the value of $\Delta$, where the $\xi_{2e}/\xi_{1e}$ ratio peaks and the $1e$ correlations are minimal. As we increase $\Delta$ and approach the FL, the Cooper-pair wave function becomes more extended (see Fig.~\ref{fig:correlations}\textbf{d,e}). Similar behavior was observed in Ref.~\cite{wangChiralSuperconductivityFractional2025} for a chiral $f$-wave superconductor. Intriguingly, in our case, $\ell_\mathrm{CP}$ also increases as we approach the FCI phase.  

Until now, we have focused on the transition being driven by a change in $\Delta$. We find that equivalently, the transition can also be driven by increasing the next-nearest-neighbor hopping $t'$, for fixed $\Delta$; see Appendix~\ref{app:paramterscans}. Even when the bandwidth is tuned independently by increasing $t'$, we observe superconducting signatures only at finite $\Delta$. This shows that increasing the bandwidth alone is insufficient and suggests that breaking the one-site magnetic translation symmetry appears to play an essential role in stabilizing superconductivity.

Although the chiral superconductor is connected to the weak-coupling regime on the metallic side, the seemingly continuous transition to the FCI at $\nu=1/3$ is clearly beyond a weak-coupling description. Hence, our model provides a controlled setting to study superconductivity in finite-size numerics, ranging from a weak-coupling SC, induced by attraction from repulsion, to the strongly-correlated regime, where the SC connects with fractional quantum Hall states.

\section{Competing phases at 1/2 filling} \label{sec:GHC}
Although we find signatures of a chiral $p$-wave superconductor over a wide range of fillings, $\nu=1/2$ is a notable exception. At this commensurate filling, we find that the system stabilizes a charge density wave. The lattice translation-symmetry breaking is visible in the sublattice-resolved structure factor
\begin{equation}
    S_{ab}(\mathbf{k})
    =
    \langle \rho_{a,-\mathbf{k}}\rho_{b,\mathbf{k}}\rangle
    -
    \langle \rho_{a,-\mathbf{k}}\rangle
    \langle \rho_{b,\mathbf{k}}\rangle ,
\end{equation}
where $\rho_{a,\mathbf{k}}$ is the Fourier-transformed density on sublattice $a$. We observe a sharp peak at $\mathbf{k}=(\pi,\pi)$ in $S_{AA}$, whereas $S_{BB}$ remains comparatively uniform; see Fig.~\ref{fig:halffilling}\textbf{a}. Thus, the density modulation resides predominantly on the less populated $A$ sublattice.

Despite this crystalline order, the charge-density-wave develops a topological response: threading one flux quantum through the cylinder pumps one unit of charge, corresponding to $\sigma_{xy}=e^2/h$; see Fig.~\ref{fig:halffilling}\textbf{c}. Correspondingly, the entanglement spectrum exhibits a chiral mode with counting $(1,1,2,\ldots)$, together with a folded replica shifted by $k_y=\pi$ because of the enlarged unit cell (Fig.~\ref{fig:halffilling}\textbf{b}). 
We refer to this state as a generalized Hall crystal (GHC), where the modifier `generalized' highlights that the state only breaks discrete translational symmetry.
This state is closely related to the re-entrant integer quantum Hall states observed experimentally~\cite{xu_SC_2025}. We further find that the GHC remains stable even for values of $\Delta>1.1t$, where the non-interacting band structure is topologically trivial (see Appendix~\ref{app:GHC}).

To better understand the GHC, we perform self-consistent Hartree-Fock calculations for our Hamiltonian Eq.~\eqref{eq:model}. Remarkably, at this level of approximation, both a trivial charge density wave with $C=0$ and the GHC with $C=1$ are closely competing and have numerically identical energies at the Hartree-Fock level. 
Which state is picked thus depends on higher-order many-body corrections. From our RPA analysis, we find that the bare nearest-neighbor interaction induces attractive next-nearest-neighbor components $V'$ upon resummation (analysis for $\nu=1/2$ see Appendix~\ref{app:RPA}), which, when inserted into our Hartree-Fock analysis, stabilize the topological GHC. Conversely, when further-range repulsion is added by hand, the trivial charge density wave with $\sigma_{xy}=0$ is energetically favored over the GHC.
The two crystals further differ in their symmetry-breaking pattern: In full agreement with our iDMRG results, the GHC in Hartree-Fock has a uniform density on the $B$ sublattice, while a CDW forms on the less populated $A$ sublattice (see Fig.~\ref{fig:halffilling}\textbf{d}). The converse applies to the trivial CDW stabilized by repulsive $V'$. Here, the minority $A$ sublattice is uniform, and the majority $B$ sublattice forms a CDW (see Fig.~\ref{fig:halffilling}\textbf{e}).   

The energetic selection favoring the GHC for $V'<0$ can already be understood from the Hartree energy. At fixed average occupation of a sublattice, a $(\pi,\pi)$ density modulation reduces the density products on next-nearest neighbors. {An attractive next-nearest-neighbor interaction, $V'<0$, which seeks to maximize the product of densities between same-sublattice sites, favors  a uniform density on the higher occupied $B$ sublattice, displacing the crystalline order to the minority $A$ sublattice.} 
A repulsive $V'>0$ favors the opposite pattern.

The Hartree-Fock analysis also offers a simple interpretation for why the GHC has a non-trivial Hall response. The self-consistently obtained state, which is at a density of $n_e = 1/4$ electrons per site of the original lattice, can be understood as an effective Lieb lattice, where every other site of the $A$ sublattice is projected out. The resulting Lieb lattice has a three-site unit cell, {with the remaining blue sites forming the sites and the red sites the bonds of the Lieb lattice (see Fig.~\ref{fig:halffilling}\textbf{d})}. In this three-site unit cell, the effective density is $n_e=1/3$ electrons per site, so that the lowest band is fully filled. The complex nearest-neighbor hopping, combined with the next-nearest-neighbor hopping inherited from the parent checkerboard model, then opens a gap and yields a Chern number of $C=1$ for the filled band. We remark that the Fock term renormalizes the effective hopping, but we find numerically that only the amplitudes change, while the phases remain invariant.

Our combined results therefore identify interaction-generated longer-range attraction as essential not only for chiral pairing, but also for stabilizing the GHC at half filling. This common microscopic origin may help explain why superconductivity and GHCs frequently appear in close proximity in experiments, where the latter have been
interpreted as re-entrant integer quantum Hall states~\cite{xu_SC_2025}.

{In the flat-band regime, our weak-coupling analysis is no longer controlled. 
The appearance of superconductivity over a range of fillings and of Jain states on both sides of half-filling motivates looking for a parent state at $\nu=1/2$ that could organize the competing phases in the phase diagram. Our numerics currently cannot resolve the ground state at half filling in the small $\Delta$ limit, hence we have to consider different possible scenarios.  
At $\nu=1/2$, it is natural to consider composite fermion states including paired states of composite fermions~\cite{readgreen_2000}. 
One candidate is the Moore-Read Pfaffian (Pf), for which non-abelian QCD$_3$ Chern-Simons theory has been proposed for a direct transition into a $p-ip$ superconductor at fixed filling, which also generalizes to finite doping~\cite{Ma2020,shi2025}. 
A simpler QED$_3$ transition to the $p-ip$ superconductor can be achieved for another  candidate topological order~\cite{shi2025}, the Particle-Hole Pfaffian (PH-Pfaffian or T-Pfaffian)~\cite{TPfaffian,CTPfaffian,Son_2015}.   
With these approaches, a connection to the $p-ip$ superconductor, which occupies a large portion of our phase diagram, could be established. 
Starting from a GHC, a direct transition to the anti-Pfaffian is possible as well~\cite{Ma2020}. It remains an interesting open question to analyze whether a unified field-theoretic description of transitions between all the competing phases can exist.
Variational Monte Carlo studies~\cite{zhang2011topological, lotrivc2026paired}, which have access to larger system sizes, could test this organizing principle and also evaluate the energetics of other candidate parent states. }

\section{Outlook \& Discussions} 
\label{sec:Outlook}

In this work, we have studied the instabilities of repulsively interacting electrons in a partially filled topological band. 
{We find that within the random phase approximation, effective attraction is generated on next-nearest-neighbor sites, which favors the formation of chiral $p$-wave superconductivity at generic fillings.}
At $\nu=1/2$, where the electron density is commensurate with the underlying square lattice, the same interactions that favor superconductivity stabilize a generalized Hall crystal.

Remarkably, we also observe both the predicted chiral superconductor and the generalized Hall crystal in unbiased numerics of finite systems. This strong tendency for the system to superconduct, along with the matching phases, highlights the favorable conditions provided by the non-trivial quantum geometry and proximity to a van Hove singularity present in our model. Importantly, signatures of superconductivity are most robust in the strongly correlated regime, and disappear once fractional quantum Hall states at Jain fillings form. In this regime, ordinary weak-coupling descriptions break down, and our tensor-network simulations find bandwidth-tuned transitions from chiral superconductivity to the fractional quantum Hall effect.
{In cylinders of size $L_y=8$, the growth of the charge-$1e$ correlation length with bond dimension at the transition between FCI and SC, and a smooth first derivative of the ground state energy, are consistent with a continuous transition where the electron becomes gapless. Smooth energy curves are also seen for other Jain fillings crossing over into the superconductor, although further analysis at these fractions and scaling with cylinder circumference is needed to assess continuity in the 2D limit. }
{We remark that, intriguingly, signatures for superconductivity at the Jain fillings are already present at the smallest possible iDMRG unit cell size. Conversely, for all other fillings away from $\nu=1/2$, superconductivity only appears when we allow unit cells larger than the minimal size (see Appendix~\ref{app:fillingPhaseDiagram}). Whether this is a purely numerical curiosity or has physical implications, we leave for future studies. In any case, we stress that capturing superconductivity within iDMRG requires careful consideration of different unit cells.}

Our model captures several features of twisted MoTe$_2$ bilayers~\cite{xu_SC_2025}. The two sublattices mimic the layer-polarized orbitals of the moir\'e lattice, and the staggered potential $\Delta$ is analogous to a displacement field, which controls the dispersion and density of states. Similar to our theoretical phase diagram, recent experiments have reported superconductivity, FCIs, and re-entrant integer quantum Hall states in a single device~\cite{xu_SC_2025}, while GHCs at $\nu=1/2$ were recently 
predicted~\cite{Sheng2024QAHC} and observed~\cite{liu2025fractionalcherninsulatorstopological, wu2026observationreconstructedcherninsulator}, which further hints at close connections between these seemingly distinct phases~\cite{lotrič2026fractionalcherninsulatorsuperconductortransition}.

Since the superconducting signatures are most robust at intermediate to strong coupling, the most important outstanding question is finding an appropriate strong-coupling mechanism for our observed superconductor. Critical points of the SC-FCI transition can provide important organizing principles. While various possible theories have been put forward~\cite{Ma2020,pichlerMicroscopicMechanismAnyon,shiAnyonDelocalizationTransitions2025, shiNonAbelianTopologicalSuperconductivity2025}, existing proposals do not capture transitions from generic Jain-states to our $p-ip$ SC. This leaves a better theoretical understanding of such transitions as an interesting direction for future work. 
In addition, larger-scale numerical simulations could help identify the fate of FCIs at $1/2$ filling, where proximity to chiral SC could favor nonabelian topological states.

More broadly, our results show that attraction generated from repulsion can provide a common organizing principle for superconductivity and generalized Hall crystallization, which compete with fractionalized states with topological order in strongly interacting Chern bands.

\textit{\textbf{Acknowledgments.---}} We thank Junkai Dong and Zijian Wang for useful discussions. F.P. and M.K. acknowledge support from the Deutsche Forschungsgemeinschaft (DFG, German Research Foundation) under Germany’s Excellence Strategy--EXC--2111--390814868, TRR 360 – 492547816 and DFG grants No. KN1254/1-2, KN1254/2-1, the European Union (grant agreement No. 101169765), as well as the Munich Quantum Valley (MQV), which is supported by the Bavarian state government with funds from the Hightech Agenda Bayern Plus. 

\textit{\textbf{Data availability.---}}Data, data analysis, and simulation codes are available upon reasonable request on Zenodo~\cite{zenodo}.

\addtocontents{toc}{\string\tocdepth@munge}

\renewcommand{\thefigure}{A\arabic{figure}}

\renewcommand{\theHfigure}{A\arabic{figure}}
\setcounter{figure}{0}

\appendix

\section{Characterizing the FCI phases} \label{app:FCI}

\begin{figure*}
\begin{center}
\includegraphics[width=0.99\linewidth]{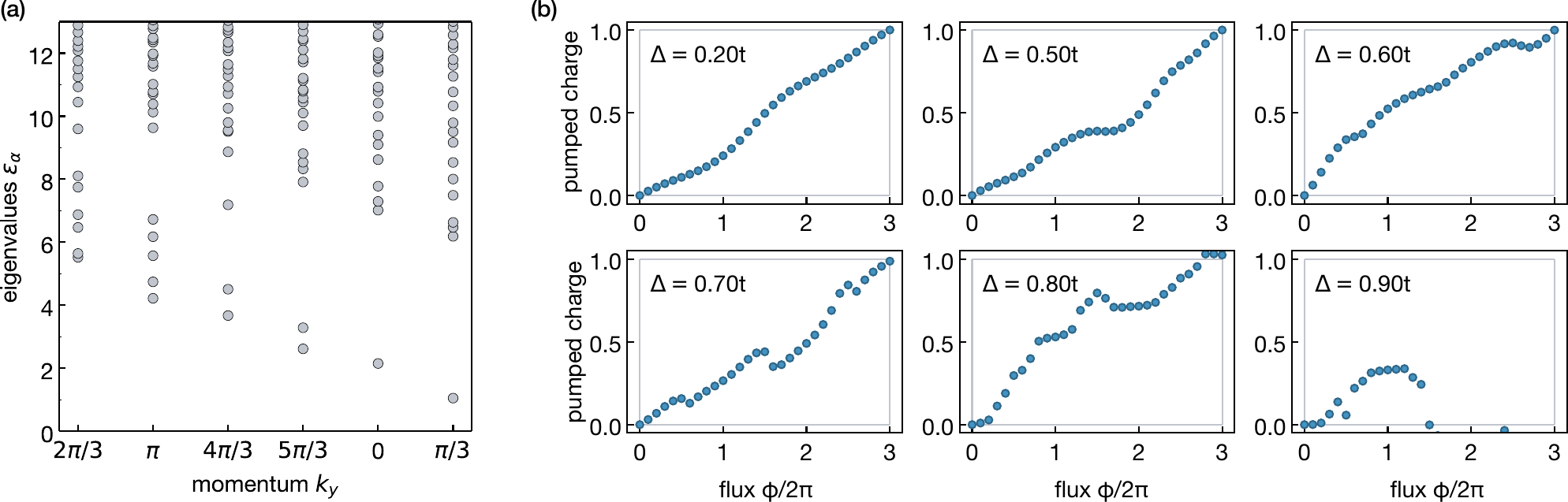}
\caption{\textbf{Fractional Chern insulator at $\nu=1/3$.} \textbf{(a)} Entanglement spectrum at $\nu=1/3$ for $L_y=12$ corresponding to $N_y^{\text{\tiny UC}} = L_y/2 = 6$ unit cells (and six conserved momenta in $y$ direction, accordingly). Here, the system is deep inside the FCI phase, with $\Delta=0.2t$, $V=2.2t$, and bond dimension $\chi=2048$. \textbf{(b)} Charge pumping for different $\Delta$ at $V=2.2t$ and $L_y = 8$, $\chi=512$. Inside the FCI, one quantized charge is pumped after inserting three flux quanta. The quantized charge pumping breaks down for $\Delta \gtrsim 0.8t$.}
\label{fig:FCI_1third}
\end{center}
\end{figure*}

To identify FCI phases, we use the entanglement spectrum and charge pumping under adiabatic flux insertion. We observe the most stable FCI at $\nu=1/3$ for small $\Delta$, where a clear chiral mode with the characteristic degeneracy counting $(1, 1, 2, 3, 5, ...)$ is visible in the entanglement spectrum (see Fig.~\ref{fig:FCI_1third}\textbf{a}). Furthermore, we find that one quantized charge is pumped along the cylinder, after we insert three flux quanta (quasi-)adiabatically, in agreement with a $\nu=1/3$ Laughlin state. Our charge-pumping protocol is as follows: First, we use iDMRG to converge to the ground state with zero flux. We then use the converged ground state as the initial state for the next run, where we insert a finite flux $\Delta \phi$ into the cylinder by shifting the single-particle momenta $k_y \rightarrow k_y + \Delta\phi$. We then let the iDMRG algorithm converge again and repeat until the desired total amount of flux was inserted. At each step, we measure the total charge at a given bond of the iMPS, thus tracking how much charge is pumped as we insert flux. This protocol mimics adiabatic flux insertion, provided that $\Delta \phi$ is sufficiently small. 

\begin{figure*}
\begin{center}
\includegraphics[width=0.99\linewidth]{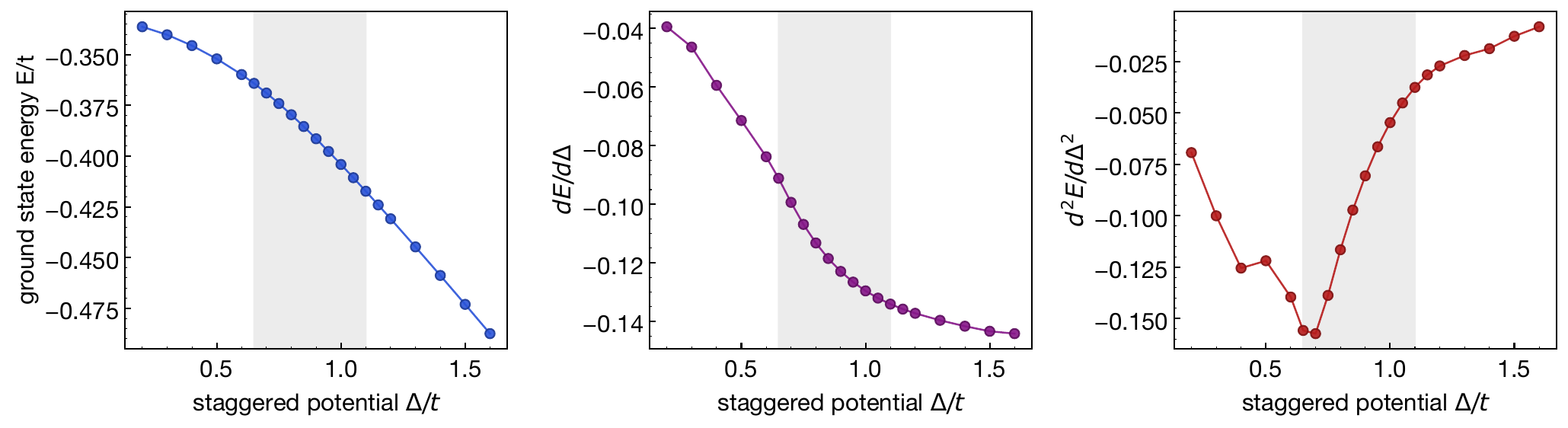}
\caption{\textbf{Ground state energy and its derivatives.} From left to right: The ground state energy (per site), its first numerical derivative, and its second numerical derivative with respect to $\Delta$, obtained from iDMRG for $L_y=8$, $\chi=4096$, $\nu=1/3$, $V=2.2t$. The shaded region indicates the putative range of superconductivity (same as in Fig.~\ref{fig:correlations}).}
\label{fig:energies_and_derivatives}
\end{center}
\end{figure*}

We find quantized charge pumping up until $\Delta \gtrsim 0.8t$, after which it breaks down (see Fig.~\ref{fig:FCI_1third}\textbf{b}). Note that the $\xi_{2e}/\xi_{1e}$ correlation-length ratio already starts to grow rapidly at $\Delta \approx 0.6t$, consistent with a continuous change from the FCI into the SC at $\nu=1/3$. This is also consistent with the ground state energy and its derivatives as a function of $\Delta$ (see Fig.~\ref{fig:energies_and_derivatives}), where only the second derivative shows non-analytic behavior, while the first derivative is still a continuous function of $\Delta$. 

\begin{figure*}
\begin{center}
\includegraphics[width=0.99\linewidth]{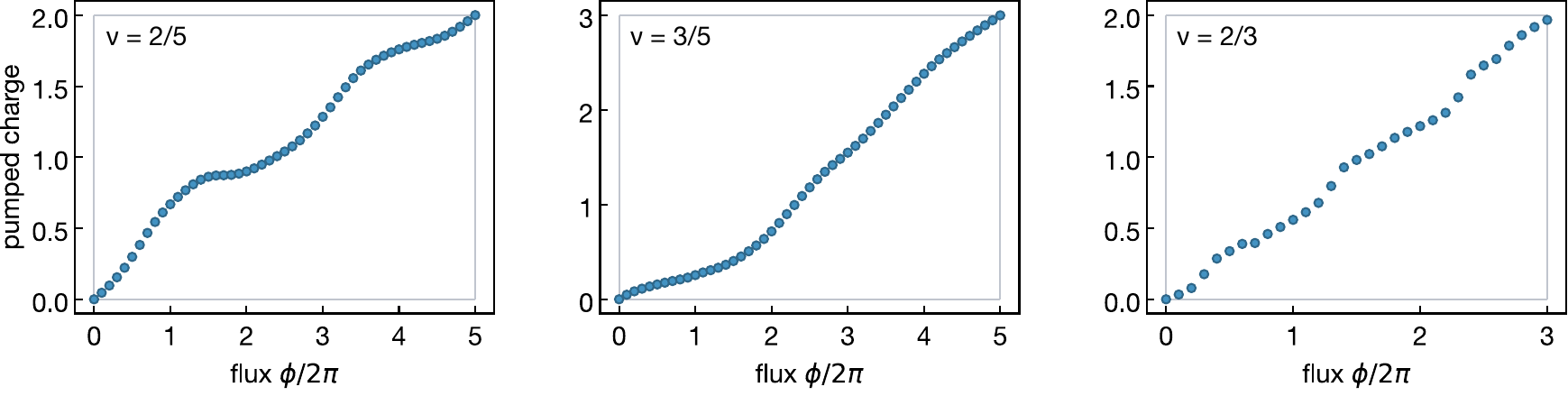}
\caption{\textbf{Charge pumping for the Jain states.} Besides the FCI at $\nu=1/3$, we also find FCIs at the Jain fillings $\nu \in \{2/5, \; 3/5, \; 2/3\}$. Here we show the pumped charge after flux insertion into the cylinder. We used $\Delta=0.5t$ and $V=2.2t$, with cylinder circumference $L_y = 8$ and bond dimension $\chi=512$.}
\label{fig:FCI_Jain}
\end{center}
\end{figure*}

We observe quantized charge pumping not only at $\nu=1/3$ but also at several fractions of the Jain sequence $\nu \in \{2/5, \; 3/5, \; 2/3\}$ (see Fig.~\ref{fig:FCI_Jain}).

\section{Filling dependent data} \label{app:fillingPhaseDiagram}

\begin{figure*}
\begin{center}
\includegraphics[width=0.5\linewidth]{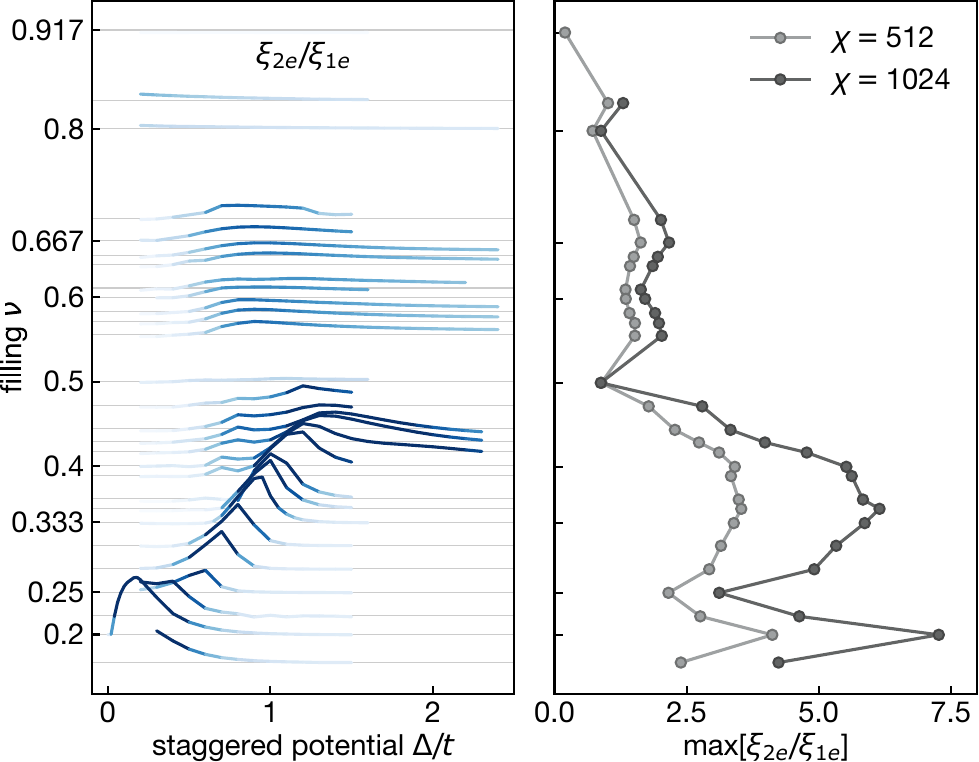}
\caption{\textbf{Filling dependent phase diagram.} Left panel: The ratio $\xi_{2e}/\xi_{1e}$ as function of staggered potential $\Delta$ for different fillings at bond dimension $\chi=1024$. Right panel: The maximum value of the $\xi_{2e}/\xi_{1e}$ taken over all $\Delta$ as a function of filling $\nu$ for two bond dimensions. The ratio increases with bond dimension for most fillings, indicating a gapped charge $1e$ sector, but a gapless charge $2e$ sector. An important exception is $\nu=1/2$, where the ratio is small and already converged in bond dimension. }
\label{fig:FillingPhaseDiagram}
\end{center}
\end{figure*}

The data underlying the phase diagram presented in the main text Fig.~\ref{fig:schematicPD} is shown in Fig.~\ref{fig:FillingPhaseDiagram}. We performed an extensive filling factor scan of the lower Chern band for $L_y = 8$, extracting $\xi_{2e}/\xi_{1e}$. For each filling, we choose a unit cell size along the cylinder direction commensurate with that filling. Furthermore, we checked several unit cell sizes for each filling to make sure our results were independent of the chosen unit cell. 
We found that for some fillings, the unit cells have to be chosen larger than the smallest possible one to obtain converged results. Remarkably, for the Jain fillings $\nu \in \{1/3, \, 2/5, \, 3/7, \, 3/5, \, 2/3 \}$ the smallest possible unit cell is already sufficient to capture the superconductor. For other fillings in between, a larger-than-minimal unit cell is required. For $\nu=1/2$, we have tried several different unit cells for a fixed cylinder circumference ($L_x=3, 6, 18$ for $L_y=8)$, all yielding the same result, i.e., no superconductor, but a GHC instead.

\section{Sublattice dependence of Cooper-pair wave function}
\label{app:Cooperpair_wavefunction}
\begin{figure*}
\begin{center}
\includegraphics[width=0.66\linewidth]{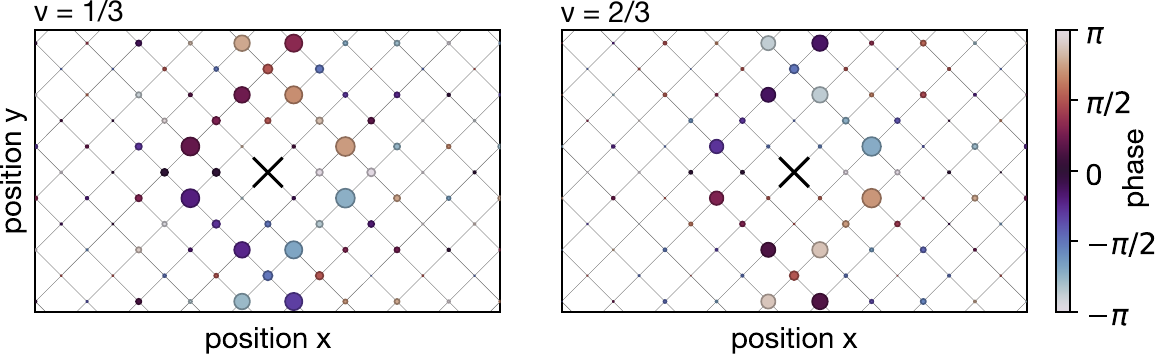}
\caption{\textbf{Cooper-pair wave function centered on $A$-sublattice site.} Real-space pairing function Eq.~\eqref{eq:pairwavefunction} from iDMRG for $\nu=1/3$ (left) and $\nu=2/3$ (right). In both cases, the origin is placed on an $A$ sublattice site, and $\Delta=0.9t$, $V=2.2t$ were used (complementing Fig.~\ref{fig:pairing_WF_DMRG}\textbf{a}, where the $B$-centered pairing wave function is shown).  }
\label{fig:appendix_pairingWF}
\end{center}
\end{figure*}

Since a finite staggered potential explicitly breaks the sublattice symmetry exchanging $A$ and $B$ sites, the real-space Cooper-pair wave function Eq.~\eqref{eq:pairwavefunction} depends on whether the origin is placed on an $A$ or $B$ site.
Given that the staggered potential $\Delta>0$ energetically favors $B$-sublattice sites, the pairing amplitude is stronger with the origin of the Cooper-pair wave function placed on a $B$ site (that is, $\Psi_{aB}(\mathbf{x})$, which is shown in the main text Fig.~\ref{fig:pairing_WF_DMRG}).
The wave function with the origin placed on an $A$ site, $\Psi_{aA}(\mathbf{x})$, is shown in Fig.~\ref{fig:appendix_pairingWF} for two representative fillings ($\nu=1/3$ and $\nu=2/3$). As expected, we find that the inter-sublattice amplitudes are the same as for the $B$-centered wave function ($|\Psi_{AB}(\mathbf{x})| = |\Psi_{BA}(\mathbf{x})|$), while the intra-sublattice amplitudes are strongly suppressed ($|\Psi_{AA}(\mathbf{x})| \ll |\Psi_{BB}(\mathbf{x})|$).

\section{Further parameter scans} \label{app:paramterscans}
\begin{figure*}
\begin{center}
\includegraphics[width=0.9\linewidth]{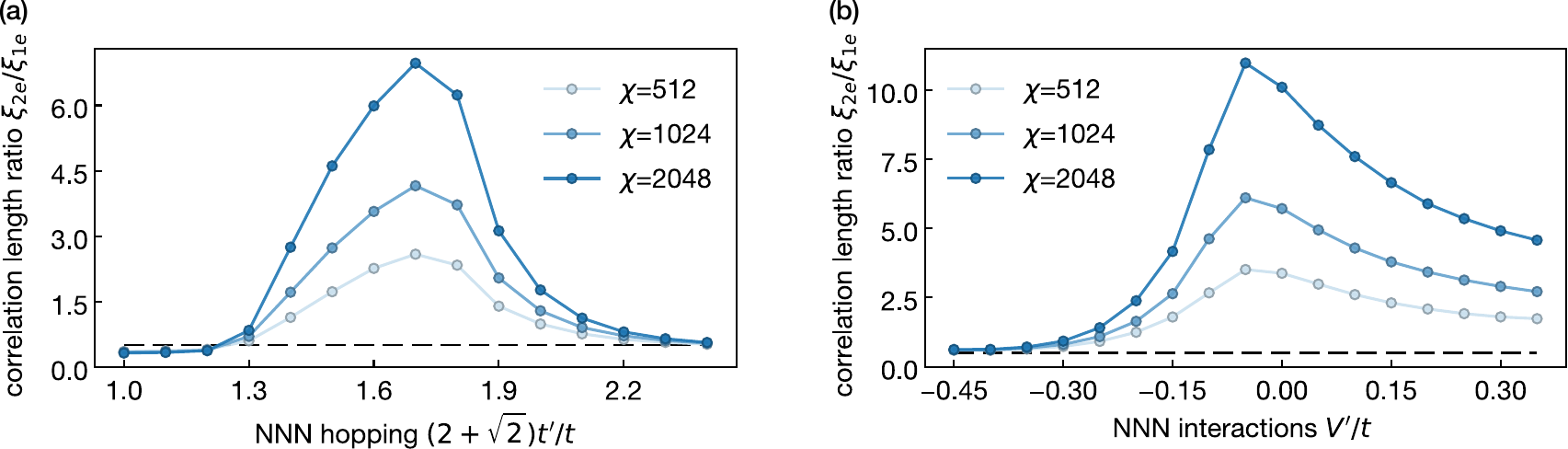}
\caption{\textbf{Further parameter scans.} \textbf{(a)} Correlation length ratio $\xi_{2e}/\xi_{1e}$ as a function of the next-nearest neighbor hopping $t'$, tuned away from the value $t'=t/(2+\sqrt{2})$ used in the main text. We fixed $\Delta=0.4t$ and $V=2.2t$. \textbf{(b)} Correlation length ratio $\xi_{2e}/\xi_{1e}$ as a function of next-nearest neighbor interactions $V'$, for $\Delta=0.9t$ and $V=2.2t$. The dashed lines correspond to $\xi_{2e}/\xi_{1e}=1/2$. }
\label{fig:otherparamters}
\end{center}
\end{figure*}

As mentioned in Sec.~\ref{sec:onethird} of the main text, the transition at $\nu=1/3$ from FCI to SC cannot only be driven by a change in $\Delta$, but also by an increasing next-nearest-neighbor hopping $t'$, which similarly increases the bandwidth. We find that for a fixed small $\Delta$, increasing $t'$ leads to a transition into an intermediate SC phase, before reaching a Fermi liquid for large $t'$ (see Fig.~\ref{fig:otherparamters}\textbf{a}). 
However, a small but finite $\Delta$ is required to break the one-site magnetic translation symmetry and make the flux-free two-site $A$--$B$ cell the primitive unit cell.
Our results, therefore, suggest that a flux-trivial primitive unit cell is an important ingredient for superconductivity in this model.

We have also tested the stability of the SC phase against bare longer-range interactions. Concretely, we find that the phase is stable against small next-nearest-neighbor interactions $V'$ (both positive and negative), see Fig.~\ref{fig:otherparamters}\textbf{b}. At fixed $V=2.2t$ and $\Delta=0.9t$, increasing the next-nearest-neighbor interaction beyond $V'=0.5t$ eventually leads to an instability towards a CDW, which we characterize by large, finite-momentum density correlation lengths. 

\section{Generalized Hall crystal from topologically trivial band}
\label{app:GHC}

\begin{figure*}
\begin{center}
\includegraphics[width=0.83\linewidth]{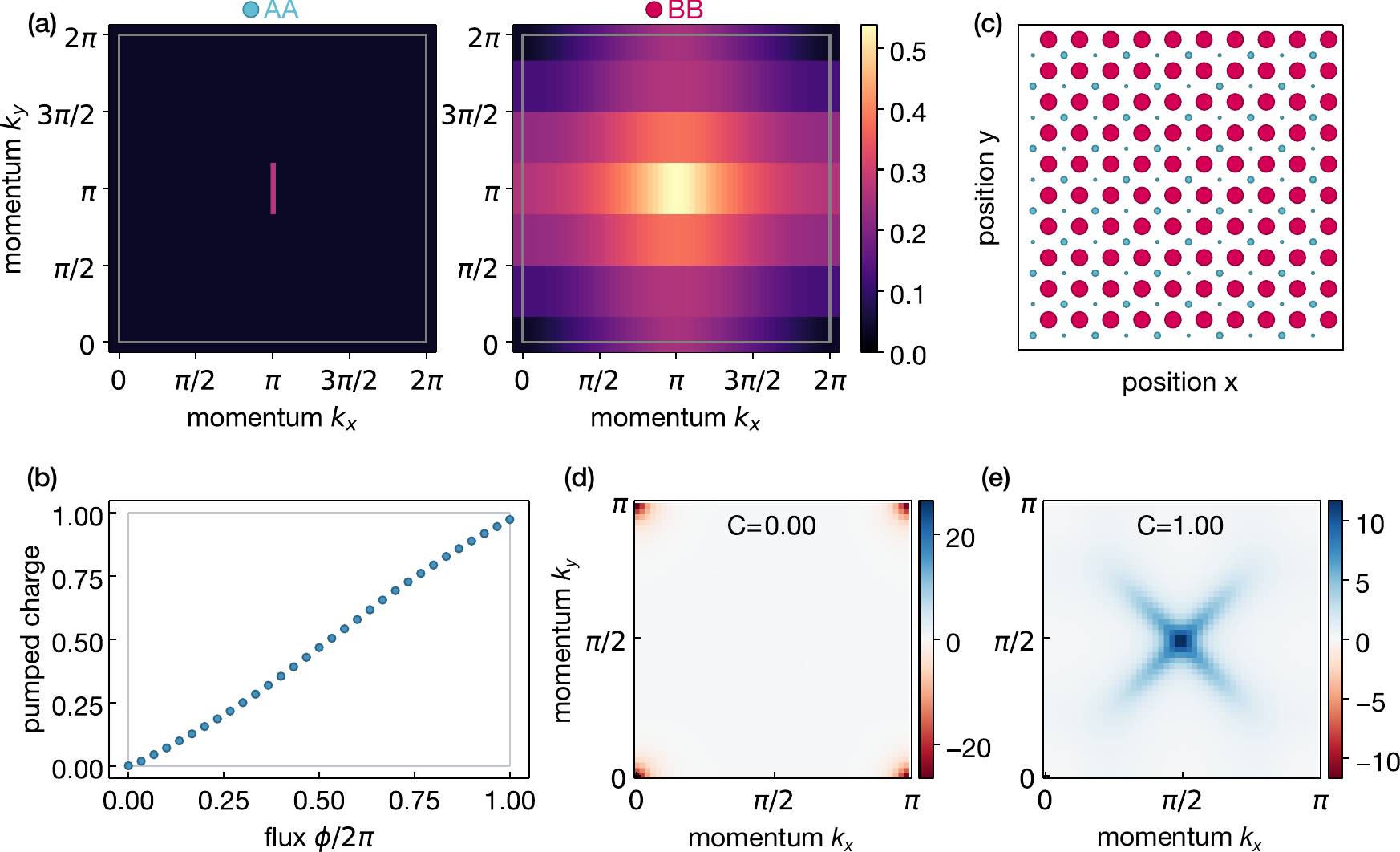}
\caption{\textbf{Generalized Hall crystal from trivial parent band.} Both iDMRG \textbf{(a, b)} and Hartree-Fock results \textbf{(c, e)} are in a regime where the single-particle bands are trivial ($\Delta>1.1t$). \textbf{(a)} Sublattice-resolved structure factor obtained from the iDMRG ground state for $L_y=12$, $\Delta=1.6t$ and $V=2.2t$ at $\nu=1/2$. \textbf{(b)} Charge pumping: when inserting one flux quantum, a single charge is pumped. \textbf{(c)} Real-space density from self-consistent Hartree-Fock calculations at $\Delta=1.4t$ with repulsive nearest-neighbor interactions $V=2.2t$ and attractive next-nearest-neighbor interactions $V'= - 0.04t$. \textbf{(d)} Berry curvature distribution of the lower single-particle band at $\Delta=1.4t$, summing to a trivial Chern number of $C=0$. In contrast, the Berry-curvature distribution of the filled Hartree-Fock band, shown in panel \textbf{(e)}, integrates to an effective Chern number $C=1$ of the GHC.}
\label{fig:appendix_GHC}
\end{center}
\end{figure*}

The GHC discussed in Sec.~\ref{sec:GHC} is stable for values of $\Delta>1.1t$, where the single-particle bands have become trivial, i.e., all single-particle bands have Chern number $C=0$ (see Fig.~\ref{fig:appendix_GHC}). Hence, its non-vanishing Hall response of $\sigma_{xy}=e^2/h$, which we confirm both from charge-pumping in iDMRG (Fig.~\ref{fig:appendix_GHC}\textbf{b}) and from the Chern number of the filled Hartree-Fock bands (Fig.~\ref{fig:appendix_GHC}\textbf{e}), is an interaction-induced effect that is not directly `inherited' from the single-particle bands. Since this state emerges from a trivial band, one may even refer to it as a generalized \emph{anomalous} Hall crystal.

\section{Renormalized interactions from RPA}
\label{app:RPA}
\begin{figure*}
\begin{center}
\includegraphics[width=0.66\linewidth]{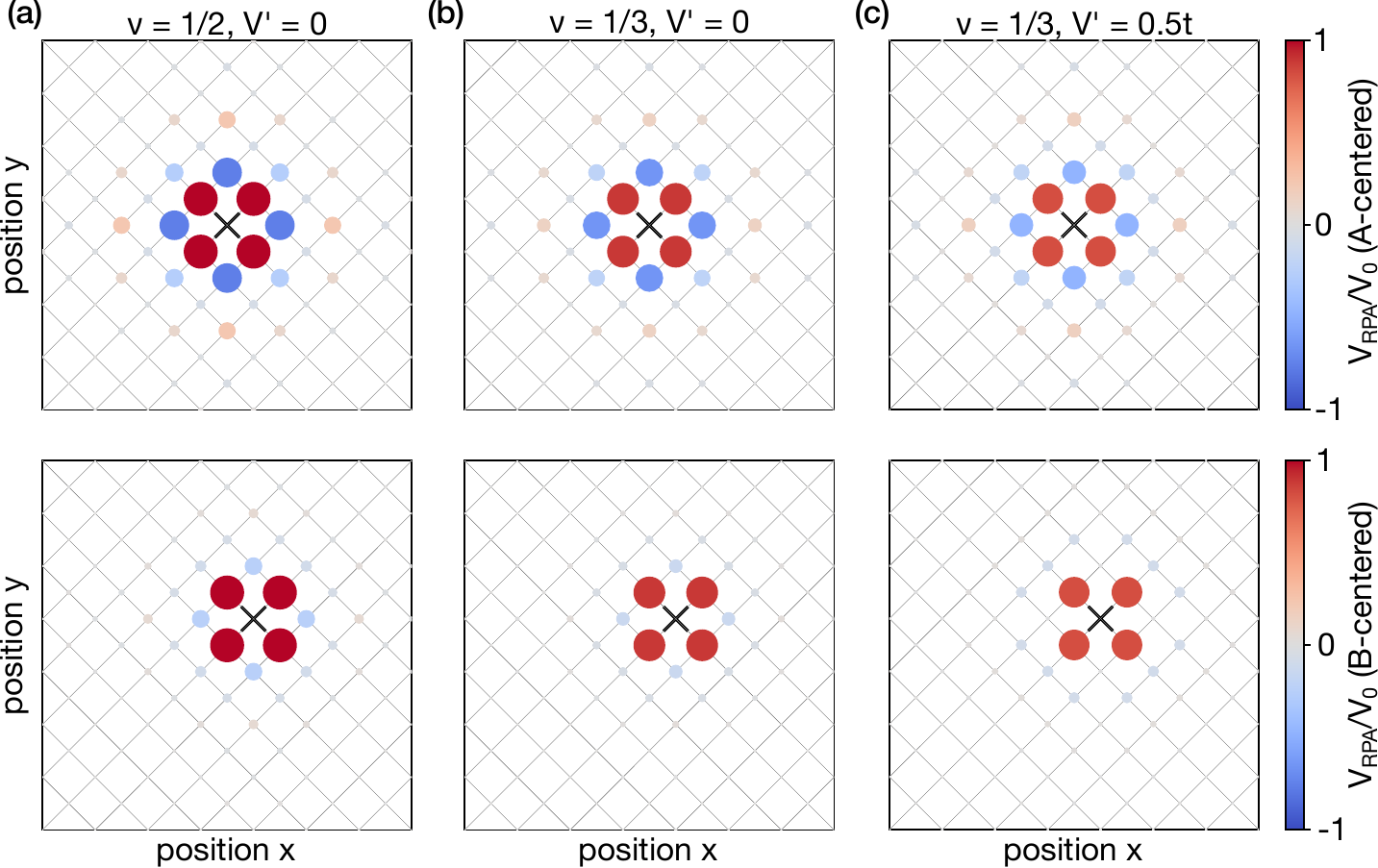}
\caption{\textbf{RPA renormalized interactions.} Effective interactions calculated using Eq.~\eqref{eq:RPA_interaction} and Fourier-transformed back into real space. The top (bottom) panels show interactions with the origin placed on an $A$ ($B$) site. Bare repulsive nearest-neighbor interactions $V=2.2t$ and a staggered potential $\Delta=0.8t$ were used. \textbf{(a)} $\nu=1/2$ and no bare next-nearest neighbor interactions $V'=0$. \textbf{(b)} $\nu=1/3$ with $V'=0$. \textbf{(c)} $\nu=1/3$ with a finite repulsive $V'=0.5t$. The effective next-nearest-neighbor interactions are attractive in all cases. }
\label{fig:appendix_RPA}
\end{center}
\end{figure*}

Effective attractive next-nearest-neighbor interactions $V'_{\mathrm{eff}}$ obtained from the random phase approximation Eq.~\eqref{eq:RPA_interaction} are a robust feature in our model, which we observe over a wide filling regime (in particular also at $\nu=1/2$, see Fig.~\ref{fig:appendix_RPA}\textbf{a},  where they stabilize the GHC, as discussed in Sec.~\ref{sec:GHC}). Furthermore, we find that effective next-nearest-neighbor attraction $V'_\mathrm{eff}<0$ persists even when we introduce a small bare repulsive next-nearest-neighbor interaction $V'>0$ (see Fig.~\ref{fig:appendix_RPA}\textbf{b},\textbf{c}). Since a finite staggered potential $\Delta$ breaks the $A-B$ sublattice symmetry, the RPA renormalized interactions break this symmetry as well, and their real-space structure depends on whether the origin is placed on an $A$ or $B$ site. We find generically stronger next-nearest-neighbor attraction with the origin placed on $A$ sites (see Fig.~\ref{fig:appendix_RPA}). 

\bibliography{library}

@article{divic_anyonsuperconductivity_2024,
author = {Stefan Divic  and Valentin Crépel  and Tomohiro Soejima  and Xue-Yang Song  and Andrew J. Millis  and Michael P. Zaletel  and Ashvin Vishwanath },
title = {{Anyon superconductivity from topological criticality in a Hofstadter–Hubbard model}},
journal = {Proceedings of the National Academy of Sciences},
volume = {122},
number = {33},
pages = {e2426680122},
year = {2025},
doi = {10.1073/pnas.2426680122},
}

@article{Shi2024,
  title = {{Doping a Fractional Quantum Anomalous Hall Insulator}},
  author = {Shi, Zhengyan Darius and Senthil, T.},
  journal = {Phys. Rev. X},
  volume = {15},
  issue = {3},
  pages = {031069},
  numpages = {31},
  year = {2025},
  month = {Sep},
  publisher = {American Physical Society},
  doi = {10.1103/kcm5-hx56},
  url = {https://link.aps.org/doi/10.1103/kcm5-hx56}
}

@article{shi2025,
	title = {{Doping lattice non-Abelian quantum Hall states}},
	pages = {150},
	author = {Shi, Zhengyan Darius and Zhang, Carolyn and Todadri, Senthil},
	journal = {SciPost Phys.},
	volume = {19},
	year = {2025},
	publisher = {SciPost},
	doi = {10.21468/SciPostPhys.19.6.150},
	url = {https://scipost.org/10.21468/SciPostPhys.19.6.150}
}

@article{Zhang2025,
  title = {Holon metal, charge density wave metal, and chiral superconductor from doping a fractional {Chern insulator and an ${\mathrm{SU}(3)}_{1}$ chiral spin liquid}},
  author = {Zhang, Ya-Hui},
  journal = {Phys. Rev. B},
  volume = {113},
  issue = {15},
  pages = {155110},
  numpages = {12},
  year = {2026},
  month = {Apr},
  publisher = {American Physical Society},
  doi = {10.1103/c8p4-637v},
  url = {https://link.aps.org/doi/10.1103/c8p4-637v}
}

@article{LongJuSc2025,
  title = {Signatures of Chiral Superconductivity in Rhombohedral Graphene},
  author = {Han, Tonghang and Lu, Zhengguang and Hadjri, Zach and Shi, Lihan and Wu, Zhenghan and Xu, Wei and Yao, Yuxuan and Cotten, Armel A. and Sharifi Sedeh, Omid and Weldeyesus, Henok and Yang, Jixiang and Seo, Junseok and Ye, Shenyong and Zhou, Muyang and Liu, Haoyang and Shi, Gang and Hua, Zhenqi and Watanabe, Kenji and Taniguchi, Takashi and Xiong, Peng and Zumb{\"u}hl, Dominik M. and Fu, Liang and Ju, Long},
  year = 2025,
  month = jul,
  journal = {Nature},
  volume = {643},
  number = {8072},
  pages = {654--661},
  doi = {10.1038/s41586-025-09169-7}
}

@article{Sahay23,
  title = {Superconductivity in a topological lattice model with strong repulsion},
  author = {Sahay, Rahul and Divic, Stefan and Parker, Daniel E. and Soejima, Tomohiro and Anand, Sajant and Hauschild, Johannes and Aidelsburger, Monika and Vishwanath, Ashvin and Chatterjee, Shubhayu and Yao, Norman Y. and Zaletel, Michael P.},
  journal = {Phys. Rev. B},
  volume = {110},
  issue = {19},
  pages = {195126},
  numpages = {26},
  year = {2024},
  month = {Nov},
  publisher = {American Physical Society},
  doi = {10.1103/PhysRevB.110.195126},
  url = {https://link.aps.org/doi/10.1103/PhysRevB.110.195126}
}

@misc{xu_SC_2025,
      title={Signatures of unconventional superconductivity near reentrant and fractional quantum anomalous {Hall} insulators}, 
      author={Fan Xu and Zheng Sun and Jiayi Li and Ce Zheng and Cheng Xu and Jingjing Gao and Tongtong Jia and Yanfei Su and Kenji Watanabe and Takashi Taniguchi and Bingbing Tong and Li Lu and Jinfeng Jia and Zhiwen Shi and Shengwei Jiang and Junhao Lin and Yuanbo Zhang and Yang Zhang and Shiming Lei and Xiaoxue Liu and Tingxin Li},
      year={2026},
      eprint={2504.06972},
      archivePrefix={arXiv},
      primaryClass={cond-mat.mes-hall},
      url={https://arxiv.org/abs/2504.06972}, 
}

@article{Xu2025_SC_theory,
  title = {{Chiral Superconductivity from Spin Polarized Chern Band in Twisted ${\mathrm{MoTe}}_{2}$}},
  author = {Xu, Cheng and Zou, Nianlong and Peshcherenko, Nikolai and Jahin, Ammar and Li, Tingxin and Lin, Shi-Zeng and Zhang, Yang},
  journal = {Phys. Rev. Lett.},
  volume = {135},
  issue = {26},
  pages = {266005},
  numpages = {7},
  year = {2025},
  month = {Dec},
  publisher = {American Physical Society},
  doi = {10.1103/h22z-4hsj},
  url = {https://link.aps.org/doi/10.1103/h22z-4hsj}
}

@article{laughlinSuperconductingGroundState1988,
  title = {Superconducting {{Ground State}} of {{Noninteracting Particles Obeying Fractional Statistics}}},
  author = {Laughlin, R. B.},
  year = {1988},
  month = jun,
  journal = {Phys. Rev. Lett.},
  volume = {60},
  number = {25},
  pages = {2677--2680},
  publisher = {American Physical Society},
  doi = {10.1103/PhysRevLett.60.2677},
  urldate = {2025-05-19}
}

@article{laughlinRelationshipHighTemperatureSuperconductivity1988,
  title = {The {{Relationship Between High-Temperature Superconductivity}} and the {{Fractional Quantum Hall Effect}}},
  author = {Laughlin, R. B.},
  year = {1988},
  month = oct,
  journal = {Science},
  volume = {242},
  number = {4878},
  pages = {525--533},
  publisher = {American Association for the Advancement of Science},
  doi = {10.1126/science.242.4878.525},
  urldate = {2025-05-19}
}

@article{chenAnyonSuperconductivity1989,
  title = {On Anyon Superconductivity},
  author = {Chen, Yi-Hong and Wilczek, Frank and Witten, Edward and Halperin, Bertrand I.},
  year = {1989},
  month = jul,
  journal = {International Journal of Modern Physics B},
  volume = {03},
  number = {07},
  pages = {1001--1067},
  publisher = {World Scientific Publishing Co.},
  doi = {10.1142/S0217979289000725},
  urldate = {2025-05-19}
}

@article{fetterRandomphaseApproximationFractionalstatistics1989,
  title = {{Random-Phase Approximation in the Fractional-Statistics Gas}},
  author = {Fetter, A. L. and Hanna, C. B. and Laughlin, R. B.},
  year = {1989},
  month = may,
  journal = {Phys. Rev. B},
  volume = {39},
  number = {13},
  pages = {9679--9681},
  publisher = {American Physical Society},
  doi = {10.1103/PhysRevB.39.9679},
  urldate = {2025-05-19}
}

@article{leeAnyonSuperconductivityFractional1989,
  title = {{Anyon Superconductivity and the Fractional Quantum {{Hall}} Effect}},
  author = {Lee, Dung-Hai and Fisher, Matthew P. A.},
  year = {1989},
  month = aug,
  journal = {Phys. Rev. Lett.},
  volume = {63},
  number = {8},
  pages = {903--906},
  publisher = {American Physical Society},
  doi = {10.1103/PhysRevLett.63.903},
  urldate = {2025-05-19}
}

@article{Lee2018,
  title = {{Emergent Multi-Flavor QED$_3$ at the Plateau Transition between Fractional Chern Insulators: Applications to Graphene Heterostructures}},
  author = {Lee, Jong Yeon and Wang, Chong and Zaletel, Michael P. and Vishwanath, Ashvin and He, Yin-Chen},
  journal = {Phys. Rev. X},
  volume = {8},
  issue = {3},
  pages = {031015},
  numpages = {21},
  year = {2018},
  month = {Jul},
  publisher = {American Physical Society},
  doi = {10.1103/PhysRevX.8.031015},
  url = {https://link.aps.org/doi/10.1103/PhysRevX.8.031015}
}

@misc{zenodo,

  author       = {Pichler, Fabian and Kuhlenkamp, Clemens and Vishwanath, Ashvin and Knap, Michael},

  title        = {{Zenodo entry for: Attraction from repulsion in an interacting Chern band:  chiral superconductivity, generalized Hall crystals, and competing fractional Chern insulators}},

  month        = oct,

  year         = 2026,

  publisher    = {Zenodo},

  doi          = {10.5281/zenodo.23103181},

}

@article{ahnSuperconductivityProximateNonAbelian2025,
  title = {Superconductivity Proximate to Non-{{Abelian}} Fractional Spin {{Hall}} Insulator in Twisted Bilayer {{MoTe2}}},
  author = {Ahn, Cheong-Eung and Seo, Donghae and Lee, Gyeoul and Kim, Youngwook and Cho, Gil Young},
  year = 2026,
  month = may,
  journal = {Nature Communications},
  volume = {17},
  number = {1},
  pages = {7005},
  doi = {10.1038/s41467-026-73777-8}
}

@article{biborskiSignaturesSuperconductingPairing2025,
  title = {Signatures of Superconducting Pairing Driven by Electron-Electron Interactions in Moir\'e {{WSe}}$_2$/{{WSe}}$_2$ Homobilayer Modelled by {{Hubbard Hamiltonian}}},
  author = {Biborski, Andrzej and Zegrodnik, Micha{\l}},
  year = 2025,
  month = oct,
  journal = {Phys. Rev. B},
  volume = {112},
  number = {16},
  pages = {L161122},
  doi = {10.1103/kdpn-5g9d},
  urldate = {2025-10-29},
}

@misc{gaoTopologicalCharge2neSuperconductors2025,
  title = {Topological {{Charge-2ne Superconductors}}},
  author = {Gao, Zhi-Qiang and Wang, Yan-Qi and Yang, Hui and Wu, Congjun},
  year = 2025,
  month = dec,
  number = {arXiv:2512.21325},
  eprint = {2512.21325},
  primaryclass = {cond-mat},
  publisher = {arXiv},
  doi = {10.48550/arXiv.2512.21325},
  urldate = {2025-12-25},
  archiveprefix = {arXiv}
}

@article{guerciFractionalizationChiralTopological2025,
  title = {{From Fractionalization to Chiral Topological Superconductivity in a Flat Chern Band}},
  author = {Guerci, Daniele and Abouelkomsan, Ahmed and Fu, Liang},
  journal = {Phys. Rev. Lett.},
  volume = {135},
  issue = {18},
  pages = {186601},
  numpages = {8},
  year = {2025},
  month = {Oct},
  publisher = {American Physical Society},
  doi = {10.1103/zm39-dstj},
  url = {https://link.aps.org/doi/10.1103/zm39-dstj}
}

@article{chen_topological_2025,
  title = {{Topological Chiral Superconductivity in the Triangular-Lattice Hofstadter-Hubbard Model}},
  author = {Chen, Feng and Wang, Wen O. and Zhang, Jia-Xin and Balents, Leon and Sheng, D. N.},
  journal = {Phys. Rev. Lett.},
  volume = {136},
  issue = {8},
  pages = {086503},
  numpages = {7},
  year = {2026},
  month = {Feb},
  publisher = {American Physical Society},
  doi = {10.1103/tm9q-w5y7},
  url = {https://link.aps.org/doi/10.1103/tm9q-w5y7}
}

@misc{hanExactModelsChiral2025,
  title = {Exact Models of Chiral Flat-Band Superconductors},
  author = {Han, Zhaoyu and {Herzog-Arbeitman}, Jonah and Gao, Qiang and Khalaf, Eslam},
  year = 2025,
  month = aug,
  number = {arXiv:2508.21127},
  eprint = {2508.21127},
  primaryclass = {cond-mat},
  publisher = {arXiv},
  doi = {10.48550/arXiv.2508.21127},
  urldate = {2025-09-01},
  archiveprefix = {arXiv}
}

@article{huangApparentInconsistencyStreda2025,
  title = {{Apparent inconsistency between Streda formula and Hall conductivity in reentrant integer quantum anomalous Hall effect in twisted ${\mathrm{MoTe}}_{2}$}},
  author = {Huang, Yi and Musser, Seth and Zhu, Jihang and Chou, Yang-Zhi and Das Sarma, Sankar},
  journal = {Phys. Rev. B},
  volume = {112},
  issue = {19},
  pages = {195136},
  numpages = {11},
  year = {2025},
  month = {Nov},
  publisher = {American Physical Society},
  doi = {10.1103/ml2q-39pp},
  url = {https://link.aps.org/doi/10.1103/ml2q-39pp}
}

@article{chen2026FinitemomentumSuperconductivityChiral,
  title = {Finite-Momentum Superconductivity from Chiral Bands in Twisted {{MoTe2}}},
  author = {Chen, Yinqi and Xu, Cheng and Zhang, Yang and Schrade, Constantin},
  year = 2026,
  month = jan,
  journal = {Nature Communications},
  volume = {17},
  number = {1},
  pages = {1077},
  doi = {10.1038/s41467-025-67836-9}
}

@misc{kuhlenkampRobustSuperconductivityDoping2025,
  title = {Robust Superconductivity upon Doping Chiral Spin Liquid and {{Chern}} Insulators in a {{Hubbard-Hofstadter}} Model},
  author = {Kuhlenkamp, Clemens and Divic, Stefan and Zaletel, Michael P. and Soejima, Tomohiro and Vishwanath, Ashvin},
  year = 2025,
  month = sep,
  number = {arXiv:2509.02675},
  eprint = {2509.02675},
  primaryclass = {cond-mat},
  publisher = {arXiv},
  doi = {10.48550/arXiv.2509.02675},
  urldate = {2025-09-04},
  archiveprefix = {arXiv}
}

@article{luscherSuperconductivityChernBand2025,
  title = {Superconductivity in a {Chern} band: Effect of time-reversal symmetry breaking on superconductivity},
  author = {L\"uscher, Bernhard E. and Fischer, Mark H.},
  journal = {Phys. Rev. B},
  volume = {112},
  issue = {21},
  pages = {214513},
  numpages = {13},
  year = {2025},
  month = {Dec},
  publisher = {American Physical Society},
  doi = {10.1103/v8ty-qlk6},
  url = {https://link.aps.org/doi/10.1103/v8ty-qlk6}
}

@article{nakajimaThermodynamicsDiluteAnyon2025,
doi = {10.1088/1742-5468/ae6c18},
url = {https://doi.org/10.1088/1742-5468/ae6c18},
year = {2026},
month = {jun},
publisher = {IOP Publishing},
volume = {2026},
number = {6},
pages = {063101},
author = {Nakajima, Yuto and Mehta, Umang and Goldman, Hart},
title = {Thermodynamics of dilute anyon gases from fusion constraints},
journal = {Journal of Statistical Mechanics: Theory and Experiment},
}

@misc{niuSimulatingTriangleHofstadterHubbard2025,
  title = {Simulating Triangle {{Hofstadter-Hubbard}} Model with Fermionic Projected Entangled Simplex States},
  author = {Niu, Sen and Sheng, D. N. and Peng, Yang},
  year = 2025,
  month = dec,
  number = {arXiv:2512.21503},
  eprint = {2512.21503},
  primaryclass = {cond-mat},
  publisher = {arXiv},
  doi = {10.48550/arXiv.2512.21503},
  urldate = {2025-12-29},
  archiveprefix = {arXiv}
}

@misc{Kalantre2026,
      title={Fermiology and the Candidate Chiral Superconductor in Rhombohedral Tetralayer Graphene}, 
      author={Sandesh S. Kalantre and Ben H. Alexander and Julian May-Mann and Jonah Herzog-Arbeitman and Marisa Hocking and Qingrui Cao and Kenji Watanabe and Takashi Taniguchi and David Goldhaber-Gordon and Andrew J. Mannix and Trithep Devakul and Yves H. Kwan and Daniel E. Parker and Aaron Sharpe},
      year={2026},
      eprint={2606.05356},
      archivePrefix={arXiv},
      primaryClass={cond-mat.supr-con},
      url={https://arxiv.org/abs/2606.05356}, 
}

@article{nosovAnyonSuperconductivityPlateau2025,
  title = {{Anyon Superconductivity and Plateau Transitions in Doped Fractional Quantum Anomalous Hall Insulators}},
  author = {Nosov, Pavel A. and Han, Zhaoyu and Khalaf, Eslam},
  journal = {Phys. Rev. Lett.},
  volume = {136},
  issue = {10},
  pages = {106501},
  numpages = {8},
  year = {2026},
  month = {Mar},
  publisher = {American Physical Society},
  doi = {10.1103/6bgj-bfdn},
  url = {https://link.aps.org/doi/10.1103/6bgj-bfdn}
}

@article{pichlerMicroscopicMechanismAnyon,
  title = {Microscopic Mechanism of Anyon Superconductivity Emerging from Fractional {{Chern}} Insulators},
  author = {Pichler, Fabian and Kuhlenkamp, Clemens and Knap, Michael and Vishwanath, Ashvin},
  year = {2026},
  month = {Mar},
  journal = {Newton},
  volume = {2},
  number = {3},
  pages = {100340},
  publisher = {Elsevier},
  doi = {10.1016/j.newton.2025.100340},
  urldate = {2026-03-02},
  url = {https://doi.org/10.1016/j.newton.2025.100340}
}

@misc{seoUnifiedCategoricalDescription2026,
  title = {A {{Unified Categorical Description}} of {{Quantum Hall Hierarchy}} and {{Anyon Superconductivity}}},
  author = {Seo, Donghae and Lee, Taegon and Cho, Gil Young},
  year = 2026,
  month = feb,
  number = {arXiv:2602.03848},
  eprint = {2602.03848},
  primaryclass = {cond-mat},
  publisher = {arXiv},
  doi = {10.48550/arXiv.2602.03848},
  urldate = {2026-02-04},
  archiveprefix = {arXiv}
}

@article{shiAnyonDelocalizationTransitions2025,
author = {Zhengyan Darius Shi  and T. Senthil },
title = {Anyon delocalization transitions out of a disordered fractional quantum anomalous {Hall} insulator},
journal = {Proceedings of the National Academy of Sciences},
volume = {122},
number = {51},
pages = {e2520608122},
year = {2025},
doi = {10.1073/pnas.2520608122},
URL = {https://www.pnas.org/doi/abs/10.1073/pnas.2520608122},
}

@misc{shiNonAbelianTopologicalSuperconductivity2025,
  title = {Non-{{Abelian}} Topological Superconductivity from Melting {{Abelian}} Fractional {{Chern}} Insulators},
  author = {Shi, Zhengyan Darius and Senthil, T.},
  year = 2025,
  month = dec,
  number = {arXiv:2512.17996},
  eprint = {2512.17996},
  primaryclass = {cond-mat},
  publisher = {arXiv},
  doi = {10.48550/arXiv.2512.17996},
  urldate = {2025-12-23},
  archiveprefix = {arXiv}
}

@misc{vilardiSuperconductivitySpin2025,
  title = {{{SC*}} {{Superconductivity}} and {{Spin Stiffnesses}} in the {{SU}}(2) {{Gauge Theory}} of the {{Two-Dimensional Hubbard Model}}},
  author = {Vilardi, Demetrio and Bonetti, Pietro M.},
  year = 2025,
  month = nov,
  number = {arXiv:2511.03436},
  eprint = {2511.03436},
  primaryclass = {cond-mat},
  publisher = {arXiv},
  doi = {10.48550/arXiv.2511.03436},
  urldate = {2025-11-06},
  archiveprefix = {arXiv}
}

@misc{wangAnyonSuperfluidTrilayer2025,
  title = {Anyon Superfluid in Trilayer Quantum {{Hall}} Systems},
  author = {Wang, Taige and Zhang, Ya-Hui},
  year = 2025,
  month = jul,
  number = {arXiv:2508.00058},
  eprint = {2508.00058},
  primaryclass = {cond-mat},
  publisher = {arXiv},
  doi = {10.48550/arXiv.2508.00058},
  urldate = {2025-08-04},
  archiveprefix = {arXiv}
}

@misc{wangChiralSuperconductivityFractional2025,
  title = {Chiral Superconductivity near a Fractional {{Chern}} Insulator},
  author = {Wang, Taige and Zaletel, Michael P.},
  year = 2025,
  month = jul,
  number = {arXiv:2507.07921},
  eprint = {2507.07921},
  primaryclass = {cond-mat},
  publisher = {arXiv},
  doi = {10.48550/arXiv.2507.07921},
  urldate = {2025-07-11},
  archiveprefix = {arXiv}
}

@misc{wangEmergentBosonic2025,
  title = {Emergent {{QED}}$_3$ at the Bosonic {{Laughlin}} State to Superfluid Transition},
  author = {Wang, Taige and Song, Xue-Yang and Zaletel, Michael P. and Senthil, T.},
  year = 2025,
  month = jul,
  number = {arXiv:2507.07611},
  eprint = {2507.07611},
  primaryclass = {cond-mat},
  publisher = {arXiv},
  doi = {10.48550/arXiv.2507.07611},
  urldate = {2025-07-11},
  archiveprefix = {arXiv}
}

@misc{zhangCharge4eAnyonSuperconductor2025,
  title = {{Charge-4$e$ {{Anyon Superconductor}} from {{Doping}} {{SU}}$(4)_1$ Chiral Spin Liquid}},
  author = {Zhang, Lu and Zhang, Ya-Hui and Song, Xue-Yang},
  year = 2025,
  month = aug,
  number = {arXiv:2508.12370},
  eprint = {2508.12370},
  primaryclass = {cond-mat},
  publisher = {arXiv},
  doi = {10.48550/arXiv.2508.12370},
  urldate = {2025-08-19},
  archiveprefix = {arXiv}
}

@article{zhangPathwaysChiralSuperconductor2025,
  title = {Pathways from a chiral superconductor to a composite Fermi liquid},
  author = {Zhang, Yunchao and Shackleton, Leyna and Senthil, T.},
  journal = {Phys. Rev. B},
  volume = {113},
  issue = {14},
  pages = {144510},
  numpages = {12},
  year = {2026},
  month = {Apr},
  publisher = {American Physical Society},
  doi = {10.1103/txjx-s4zx},
  url = {https://link.aps.org/doi/10.1103/txjx-s4zx}
}

@article{zhouChernSimonsmatterConformalField2025,
  title = {{Chern-Simons-Matter Conformal Field Theory on the Fuzzy Sphere: Higgs Transition of the Kalmeyer-Laughlin Chiral Spin Liquid}},
  author = {Zhou, Zheng and Wang, Chong and He, Yin-Chen},
  journal = {Phys. Rev. Lett.},
  volume = {137},
  issue = {10},
  pages = {106504},
  numpages = {8},
  year = {2026},
  month = {Sep},
  publisher = {American Physical Society},
  doi = {10.1103/m64z-bvm3},
  url = {https://link.aps.org/doi/10.1103/m64z-bvm3}
}

@article{Ma2020,
  title = {{Emergent ${\mathrm{QCD}}_{3}$ quantum phase transitions of fractional Chern insulators}},
  author = {Ma, Ruochen and He, Yin-Chen},
  journal = {Phys. Rev. Res.},
  volume = {2},
  issue = {3},
  pages = {033348},
  numpages = {19},
  year = {2020},
  month = {Sep},
  publisher = {American Physical Society},
  doi = {10.1103/PhysRevResearch.2.033348},
  url = {https://link.aps.org/doi/10.1103/PhysRevResearch.2.033348}
}

@misc{guerci2026topologicalSC,
      title={Topological superconductivity with emergent vortex lattice in twisted semiconductors}, 
      author={Daniele Guerci and Ahmed Abouelkomsan and Liang Fu},
      year={2026},
      eprint={2602.15106},
      archivePrefix={arXiv},
      primaryClass={cond-mat.supr-con},
      url={https://arxiv.org/abs/2602.15106}, 
}

@article{maymann2026HowPairingMechanism,
  title = {How Pairing Mechanism Dictates Topology in Valley-Polarized Superconductors with {{Berry}} Curvature},
  author = {{May-Mann}, Julian and Helbig, Tobias and Devakul, Trithep},
  year = 2026,
  month = apr,
  journal = {npj Quantum Materials},
  volume = {11},
  number = {1},
  pages = {53},
  publisher = {Nature Publishing Group},
  doi = {10.1038/s41535-026-00878-4},
  urldate = {2026-07-27}
}

@article{jahin2026EnhancedKohnLuttinger,
  title = {Enhanced {{Kohn-Luttinger}} Superconductivity in Geometric Bands},
  author = {Jahin, Ammar and Lin, Shi-Zeng},
  year = 2026,
  month = jan,
  journal = {Phys. Rev. B},
  volume = {113},
  number = {1},
  pages = {014504},
  doi = {10.1103/gt8h-czf3},
  urldate = {2026-07-27}
}

@article{shavit2026NematicEnhancementSuperconductivity,
  title = {Nematic Enhancement of Superconductivity in Multilayer Graphene via Quantum Geometry},
  author = {Shavit, Gal},
  year = 2026,
  month = mar,
  journal = {Phys. Rev. B},
  volume = {113},
  number = {9},
  pages = {094502},
  doi = {10.1103/sw7f-4x94},
  urldate = {2026-07-27}
}

@article{shavit2025QuantumGeometricKohnLuttinger,
  title = {Quantum {{Geometric Kohn-Luttinger Superconductivity}}},
  author = {Shavit, Gal and Alicea, Jason},
  year = 2025,
  month = apr,
  journal = {Phys. Rev. Lett.},
  volume = {134},
  number = {17},
  pages = {176001},
  doi = {10.1103/PhysRevLett.134.176001},
  urldate = {2026-07-27}
}

@misc{zou2026KohnLuttingerSuperconductivity,
  title = {Kohn--{{Luttinger Superconductivity}} in {{Flat Chern Bands}}},
  author = {Zou, Nianlong and Xu, Cheng and Barlas, Yafis and Zhang, Yang},
  year = 2026,
  month = jun,
  number = {arXiv:2606.17205},
  eprint = {2606.17205},
  primaryclass = {cond-mat.mes-hall},
  publisher = {arXiv},
  doi = {10.48550/arXiv.2606.17205},
  urldate = {2026-06-17},
  archiveprefix = {arXiv}
}

@misc{lotrič2026fractionalcherninsulatorsuperconductortransition,
      title={Is the fractional {Chern} insulator-superconductor transition in twisted {MoTe$_2$} direct?}, 
      author={Tevž Lotrič and Steven H. Simon},
      year={2026},
      eprint={2609.30368},
      archivePrefix={arXiv},
      primaryClass={cond-mat.str-el},
      url={https://arxiv.org/abs/2609.30368}, 
}

@misc{lotric2026phasesitinerantanyonslaughlins,
      title={{Phases of itinerant anyons in Laughlin's quantum Hall states on a lattice}}, 
      author={Tevž Lotrič and Steven H. Simon},
      year={2026},
      eprint={2603.22389},
      archivePrefix={arXiv},
      primaryClass={cond-mat.str-el},
      url={https://arxiv.org/abs/2603.22389}, 
}

@misc{zhang2026colorsuperconductorsholonmetals,
      title={{Color superconductors and holon metals from doping a Fractional Chern insulator}}, 
      author={Ya-Hui Zhang},
      year={2026},
      eprint={2607.18238},
      archivePrefix={arXiv},
      primaryClass={cond-mat.str-el},
      url={https://arxiv.org/abs/2607.18238}, 
}

@misc{shi2026superconductivitynonfermiliquidmetals,
      title={{Superconductivity and non-Fermi liquid metals in a charge-1/3 anyon fluid}}, 
      author={Zhengyan Darius Shi and T. Senthil},
      year={2026},
      eprint={2606.20403},
      archivePrefix={arXiv},
      primaryClass={cond-mat.str-el},
      url={https://arxiv.org/abs/2606.20403}, 
}

@misc{fan2026hiddenweakpairingsuperconductivitynoninteracting,
      title={Hidden weak-pairing superconductivity of non-interacting anyons obeying $\frac{1}{3}$ statistics}, 
      author={Zheng-Duo Fan and Ashvin Vishwanath and Zijian Wang},
      year={2026},
      eprint={2605.19036},
      archivePrefix={arXiv},
      primaryClass={cond-mat.str-el},
      url={https://arxiv.org/abs/2605.19036}, 
}

@misc{wang2026topologicalsuperconductivityabelianfractional,
      title={{Topological superconductivity from Abelian fractional Chern insulators}}, 
      author={Taige Wang},
      year={2026},
      eprint={2605.29034},
      archivePrefix={arXiv},
      primaryClass={cond-mat.str-el},
      url={https://arxiv.org/abs/2605.29034}, 
}

@misc{senthil2026fractionalizedmetalsdopedanyons,
      title={Fractionalized metals from doped anyons: Application to {tMoTe2}}, 
      author={T. Senthil},
      year={2026},
      eprint={2607.11484},
      archivePrefix={arXiv},
      primaryClass={cond-mat.str-el},
      url={https://arxiv.org/abs/2607.11484}, 
}

@misc{han2026singlecomponenttwistedmathbbz3orthogonal,
      title={Single-component twisted $\mathbb{Z}_3$ orthogonal metal in an $e/3$-anyon fluid}, 
      author={Zhaoyu Han and Ashvin Vishwanath and Eslam Khalaf},
      year={2026},
      eprint={2607.15274},
      archivePrefix={arXiv},
      primaryClass={cond-mat.str-el},
      url={https://arxiv.org/abs/2607.15274}, 
}

@misc{mehta2026coloringanyonsuperconductivity,
      title={Coloring in anyon superconductivity}, 
      author={Umang Mehta and Yuto Nakajima and Hart Goldman},
      year={2026},
      eprint={2607.19470},
      archivePrefix={arXiv},
      primaryClass={cond-mat.str-el},
      url={https://arxiv.org/abs/2607.19470}, 
}

@article{Luttinger1966,
  title = {New Mechanism for Superconductivity},
  author = {Luttinger, J. M.},
  journal = {Phys. Rev.},
  volume = {150},
  issue = {1},
  pages = {202--214},
  numpages = {0},
  year = {1966},
  month = {Oct},
  publisher = {American Physical Society},
  doi = {10.1103/PhysRev.150.202},
  url = {https://link.aps.org/doi/10.1103/PhysRev.150.202}
}

@article{Kohn1965,
  title = {New Mechanism for Superconductivity},
  author = {Kohn, W. and Luttinger, J. M.},
  journal = {Phys. Rev. Lett.},
  volume = {15},
  issue = {12},
  pages = {524--526},
  numpages = {0},
  year = {1965},
  month = {Sep},
  publisher = {American Physical Society},
  doi = {10.1103/PhysRevLett.15.524},
  url = {https://link.aps.org/doi/10.1103/PhysRevLett.15.524}
}

@article{sun2011NearlyFlatbandsNontrivial,
  title = {Nearly {{Flatbands}} with {{Nontrivial Topology}}},
  author = {Sun, Kai and Gu, Zhengcheng and Katsura, Hosho and Das Sarma, S.},
  year = 2011,
  month = jun,
  journal = {Phys. Rev. Lett.},
  volume = {106},
  number = {23},
  pages = {236803},
  publisher = {American Physical Society},
  doi = {10.1103/PhysRevLett.106.236803},
  urldate = {2024-05-10}
}

@article{neupert2011,
  title = {{Fractional Quantum Hall States at Zero Magnetic Field}},
  author = {Neupert, Titus and Santos, Luiz and Chamon, Claudio and Mudry, Christopher},
  journal = {Phys. Rev. Lett.},
  volume = {106},
  issue = {23},
  pages = {236804},
  numpages = {4},
  year = {2011},
  month = {Jun},
  publisher = {American Physical Society},
  doi = {10.1103/PhysRevLett.106.236804},
  url = {https://link.aps.org/doi/10.1103/PhysRevLett.106.236804}
}

@article{Sheng2011,
	author = {Sheng, D. N. and Gu, Zheng-Cheng and Sun, Kai and Sheng, L.},
	date = {2011/07/12},
	doi = {10.1038/ncomms1380},
	id = {Sheng2011},
	isbn = {2041-1723},
	journal = {Nature Communications},
	number = {1},
	pages = {389},
	title = {Fractional quantum {Hall} effect in the absence of {Landau} levels},
	url = {https://doi.org/10.1038/ncomms1380},
	volume = {2},
	year = {2011}}

@article{regnault2011FractionalChernInsulator,
  title = {Fractional {{Chern Insulator}}},
  author = {Regnault, N. and Bernevig, B. Andrei},
  year = 2011,
  month = dec,
  journal = {Phys. Rev. X},
  volume = {1},
  number = {2},
  pages = {021014},
  publisher = {American Physical Society},
  doi = {10.1103/PhysRevX.1.021014},
  urldate = {2024-04-22}
}

@article{Hauschild2018Tenpy,
	title = {Efficient numerical simulations with Tensor Networks: Tensor Network Python (TeNPy)},
	pages = {5},
	author = {Hauschild, Johannes and Pollmann, Frank},
	journal = {SciPost Phys. Lect. Notes},
	year = {2018},
	publisher = {SciPost},
	doi = {10.21468/SciPostPhysLectNotes.5},
	url = {https://scipost.org/10.21468/SciPostPhysLectNotes.5}
}

@article{Gannot2023,
  title = {How quantum phases on cylinders approach the two-dimensional limit},
  author = {Gannot, Yuval and Kivelson, Steven A.},
  journal = {Phys. Rev. B},
  volume = {107},
  issue = {7},
  pages = {075127},
  numpages = {15},
  year = {2023},
  month = {Feb},
  publisher = {American Physical Society},
  doi = {10.1103/PhysRevB.107.075127},
  url = {https://link.aps.org/doi/10.1103/PhysRevB.107.075127}
}

@misc{hu2026quantumgeometryinduced,
      title={{Quantum-Geometry-Induced Superconductivity near a Fractional Chern Insulator}}, 
      author={Haoyu Hu and Lei Chen},
      year={2026},
      eprint={2606.21166},
      archivePrefix={arXiv},
      primaryClass={cond-mat.supr-con},
      url={https://arxiv.org/abs/2606.21166}, 
}

@article{readgreen_2000,
  title = {{Paired states of fermions in two dimensions with breaking of parity and time-reversal symmetries and the fractional quantum Hall effect}},
  author = {Read, N. and Green, Dmitry},
  journal = {Phys. Rev. B},
  volume = {61},
  issue = {15},
  pages = {10267--10297},
  numpages = {0},
  year = {2000},
  month = {Apr},
  publisher = {American Physical Society},
  doi = {10.1103/PhysRevB.61.10267},
  url = {https://link.aps.org/doi/10.1103/PhysRevB.61.10267}
}

@article{white1992DMRG,
  title = {Density matrix formulation for quantum renormalization groups},
  author = {White, Steven R.},
  journal = {Phys. Rev. Lett.},
  volume = {69},
  issue = {19},
  pages = {2863--2866},
  numpages = {0},
  year = {1992},
  month = {Nov},
  publisher = {American Physical Society},
  doi = {10.1103/PhysRevLett.69.2863},
  url = {https://link.aps.org/doi/10.1103/PhysRevLett.69.2863}
}

@misc{wu2026observationreconstructedcherninsulator,
      title={{Observation of a Reconstructed Chern Insulator in Twisted Bilayer MoTe2}}, 
      author={Min Wu and Lingxiao Li and Yunze Ouyang and Yifan Jiang and Wenxuan Qiu and Zaizhe Zhang and Zihao Huo and Qiu Yang and Ming Tian and Neng Wan and Kenji Watanabe and Takashi Taniguchi and Shiming Lei and Fengcheng Wu and Xiaobo Lu},
      year={2026},
      eprint={2603.16374},
      archivePrefix={arXiv},
      primaryClass={cond-mat.mes-hall},
      url={https://arxiv.org/abs/2603.16374}, 
}

@misc{liu2025fractionalcherninsulatorstopological,
      title={{From fractional Chern insulators to topological electronic crystals in moir\'e MoTe2: quantum geometry tuning via remote layer}}, 
      author={Feng Liu and Fan Xu and Cheng Xu and Jiayi Li and Zheng Sun and Jiayong Xiao and Ning Mao and Xumin Chang and Xinglin Tao and Kenji Watanabe and Takashi Taniguchi and Jinfeng Jia and Ruidan Zhong and Zhiwen Shi and Shiyong Wang and Guorui Chen and Xiaoxue Liu and Dong Qian and Yang Zhang and Tingxin Li and Shengwei Jiang},
      year={2025},
      eprint={2512.03622},
      archivePrefix={arXiv},
      primaryClass={cond-mat.mes-hall},
      url={https://arxiv.org/abs/2512.03622}, 
}

@article{Sheng2024QAHC,
  title = {{Quantum Anomalous Hall Crystal at Fractional Filling of Moir\'e Superlattices}},
  author = {Sheng, D. N. and Reddy, Aidan P. and Abouelkomsan, Ahmed and Bergholtz, Emil J. and Fu, Liang},
  journal = {Phys. Rev. Lett.},
  volume = {133},
  issue = {6},
  pages = {066601},
  numpages = {7},
  year = {2024},
  month = {Aug},
  publisher = {American Physical Society},
  doi = {10.1103/PhysRevLett.133.066601},
  url = {https://link.aps.org/doi/10.1103/PhysRevLett.133.066601}
}

@article{lotrivc2026paired,
  title={Paired parton trial states for the superfluid-fractional {C}hern insulator transition},
  author={Lotri{\v{c}}, Tev{\v{z}} and Simon, Steven H},
  journal={Phys. Rev. Lett.},
  volume={136},
  number={9},
  pages={096601},
  year={2026},
  publisher={APS},
  url={https://journals.aps.org/prl/abstract/10.1103/l9rp-y68y}
}

@article{TPfaffian,
  title = {{Non-Abelian Topological Order on the Surface of a 3D Topological Superconductor from an Exactly Solved Model}},
  author = {Fidkowski, Lukasz and Chen, Xie and Vishwanath, Ashvin},
  journal = {Phys. Rev. X},
  volume = {3},
  issue = {4},
  pages = {041016},
  numpages = {17},
  year = {2013},
  month = {Nov},
  publisher = {American Physical Society},
  doi = {10.1103/PhysRevX.3.041016},
  url = {https://link.aps.org/doi/10.1103/PhysRevX.3.041016}
}

@article{CTPfaffian,
  title = {{Symmetry enforced non-Abelian topological order at the surface of a topological insulator}},
  author = {Chen, Xie and Fidkowski, Lukasz and Vishwanath, Ashvin},
  journal = {Phys. Rev. B},
  volume = {89},
  issue = {16},
  pages = {165132},
  numpages = {15},
  year = {2014},
  month = {Apr},
  publisher = {American Physical Society},
  doi = {10.1103/PhysRevB.89.165132},
  url = {https://link.aps.org/doi/10.1103/PhysRevB.89.165132}
}

@article{Son_2015,
  title = {{Is the Composite Fermion a Dirac Particle?}},
  author = {Son, Dam Thanh},
  journal = {Phys. Rev. X},
  volume = {5},
  issue = {3},
  pages = {031027},
  numpages = {14},
  year = {2015},
  month = {Sep},
  publisher = {American Physical Society},
  doi = {10.1103/PhysRevX.5.031027},
  url = {https://link.aps.org/doi/10.1103/PhysRevX.5.031027}
}

@article{zhang2011topological,
  title = {Topological entanglement entropy of {${\mathbb{Z}}_{2}$} spin liquids and lattice Laughlin states},
  author = {Zhang, Yi and Grover, Tarun and Vishwanath, Ashvin},
  journal = {Phys. Rev. B},
  volume = {84},
  issue = {7},
  pages = {075128},
  numpages = {7},
  year = {2011},
  month = {Aug},
  publisher = {American Physical Society},
  doi = {10.1103/PhysRevB.84.075128},
  url = {https://link.aps.org/doi/10.1103/PhysRevB.84.075128}
}

\end{document}